\documentclass[a4paper,oneside,11pt]{article}
\usepackage{graphicx}
\usepackage{multicol}
\usepackage{color}
\definecolor{rosso}{cmyk}{0,1,1,0.4}
\definecolor{rossos}{cmyk}{0,1,1,0.55}
\definecolor{rossoc}{cmyk}{0,1,1,0.2}
\definecolor{blu}{cmyk}{1,1,0,0.3}
\definecolor{blus}{cmyk}{1,1,0,0.6}
\definecolor{bluc}{cmyk}{1,1,0,0.1}
\definecolor{verde}{cmyk}{0.92,0,0.59,0.25}
\definecolor{verdec}{cmyk}{0.92,0,0.59,0.15}
\definecolor{verdes}{cmyk}{0.92,0,0.59,0.4}
\newcommand{\beq}{\begin{equation}}
\newcommand{\eeq}{\end{equation}}
\newcommand{\diag}{\hbox{diag}\,}
\newcommand{\GeV}{\,{\rm GeV}}

\newcommand{\lambdaN}{\lambda}

\font\tenrsfs=rsfs10
\font\sevenrsfs=rsfs7
\font\fiversfs=rsfs5
\newfam\rsfsfam
\textfont\rsfsfam=\tenrsfs
\scriptfont\rsfsfam=\sevenrsfs
\scriptscriptfont\rsfsfam=\fiversfs
\def\mathscr#1{{\fam\rsfsfam\relax#1}}
\def\Lag{\mathscr{L}}

\advance\textheight by \topskip
\large\normalsize
\newcommand{\be}{\begin{equation}}
\newcommand{\ee}{\end{equation}}
\newcommand{\ba}{\begin{array}}
\newcommand{\ea}{\end{array}}

\newcommand{\eV}{\,{\rm eV}}

\def\Red  {}

\def\Black{}
\def\Blue {}
\newcommand{\eq}[1]{~(\ref{eq:#1})}

\newcommand{\NP}{Nucl. Phys.}

\newcommand{\PR}{Phys. Rev.}

\newcommand{\fig}[1]{~\ref{fig:#1}}

\def\circa#1{\,\raise.3ex\hbox{$#1$\kern-.75em\lower1ex\hbox{$\sim$}}\,}
\makeatletter
\def\art{\@ifnextchar[{\eart}{\oart}}
\def\eart[#1]#2#3#4#5#6{{\rm #2}, {#3 #4} {\rm (#6) #5} ({#1})}
\def\hepart[#1]#2{{\rm #2, #1}}
\newcommand{\oart}[5]{{\rm #1}, {#2 #3} {\rm (#5) #4}}

\newcounter{alphaequation}[equation]
\def\thealphaequation{\theequation\hbox to
0.6em{\hfil\alph{alphaequation}\hfil}}
\def\eqnsystem#1{
\def\@eqnnum{{\rm (\thealphaequation)}}
\def\@@eqncr{\let\@tempa\relax \ifcase\@eqcnt \def\@tempa{& & &} \or
   \def\@tempa{& &}\or \def\@tempa{&}\fi\@tempa
   \if@eqnsw\@eqnnum\refstepcounter{alphaequation}\fi
\global\@eqnswtrue\global\@eqcnt=0\cr}
\refstepcounter{equation} \let\@currentlabel\theequation \def\@tempb{#1}
\ifx\@tempb\empty\else\label{#1}\fi
\refstepcounter{alphaequation}
\let\@currentlabel\thealphaequation
\global\@eqnswtrue\global\@eqcnt=0 \tabskip\@centering\let\\=\@eqncr
$$\halign to \displaywidth\bgroup \@eqnsel\hskip\@centering
$\displaystyle\tabskip\z@{##}$&\global\@eqcnt\@ne
\hskip2\arraycolsep\hfil${##}$\hfil&  
\global\@eqcnt\tw@\hskip2\arraycolsep
$\displaystyle\tabskip\z@{##}$\hfil
\tabskip\@centering&\llap{##}\tabskip\z@\cr}
\def\endeqnsystem{\@@eqncr\egroup$$\global\@ignoretrue} \makeatother

\newcommand{\riga}[1]{\noalign{\hbox{\parbox{\textwidth}{#1}}}\nonumber}

\begin{document}%\twocolumn[
\centerline{hep-ph/0312203\hfill IFUP--TH/2003-48}
\Black
\vspace{1.0cm}
\centerline{\LARGE\bf\Red Constraints on neutrino masses from leptogenesis models}
\medskip\bigskip\Black\vspace{0.6cm}
   \centerline{\large\bf Thomas Hambye$^a$, Yin Lin$^b$,
Alessio Notari$^b$,}\vspace{0.15cm}
\centerline{\large\bf Michele Papucci$^b$ {\rm and} Alessandro  
Strumia$^c$}\vspace{0.5cm}
   \centerline{\em $^a$ Department of Physics, Theoretical Physics,
University of Oxford, Oxford OX1\hspace{0.2em}3NP, UK }\vspace{0.13cm}
   \centerline{\em $^b$ Scuola Normale Superiore,
Piazza dei Cavalieri 7, I--56126 Pisa and INFN, Italia}\vspace{0.13cm}
   \centerline{\em $^c$ Dipartimento di Fisica dell'Universit\`a di Pisa
   and INFN, Italia}\vspace{1.cm}

\Blue\centerline{\large\bf Abstract}
\begin{quote}
%\large
\vspace{-0.25cm}
\indent
Upper bounds on the CP asymmetry relevant for leptogenesis are
reexamined and found weaker than in previous literature, both for
hierarchical and for quasi-degenerate right-handed neutrinos. 
Successful leptogenesis implies the usual 
lower bound on right-handed neutrino masses,
and an upper bound on left-handed neutrino masses (which we
obtain to be $0.15\eV$ at  $3\sigma$)
only if right-handed neutrinos 
are assumed to be $\,$much$\,$ more hierarchical than left-handed neutrinos.
Otherwise both bounds can be considerably relaxed.
The constraint on light neutrino masses varies assuming 
different interpretations of why neutrinos should be quasi-degenerate.
With conservative assumptions, 
we find that a mild quasi-degeneracy allows neutrinos  
heavier than an eV compatibly with leptogenesis.

We also extend computations of thermal leptogenesis to an
alternative model of neutrino mass mediated by fermion triplets which  
was never
considered so far for leptogenesis.
Leptogenesis can be successful despite the
effect of gauge interactions, resulting in only
slightly stronger constraints on neutrino masses.

\Black
\end{quote}

\vspace{0.5cm}\noindent
Assuming that neutrino masses are generated
by tree level exchange of right-handed neutrinos, that
the observed baryon asymmetry is produced
via thermal leptogenesis \cite{FY}
and that right-handed neutrinos are hierarchical,
one can derive interesting
constraints~\cite{leptogenesisBounds,epsP,thermal}:
right-handed neutrinos must be heavier than about $10^{8}\GeV$ and
left-handed neutrinos must be lighter than about
$0.1\eV$~\cite{epsP,thermal}.
Since the former constraint leads to a possible conflict
between leptogenesis and gravitino overproduction, and since the later
one is stronger than present experimental bounds and is
close to the mass scale suggested
by atmospheric oscillations (i.e.~$m_3 \circa{>} 0.05 \eV$), in
this article we reconsider these constraints in details.
To this end we adopt the results
of \cite{thermal} for a
precise computation of the dynamics of
thermal leptogenesis, and we reexamine the upper bound on
CP violation in right-handed neutrino decays.

In section~\ref{H} we consider a hierarchical spectrum of
right-handed neutrinos. We show that the lower bound above on
their masses can be significantly evaded dropping
the assumption (made in~\cite{leptogenesisBounds}) that
the hierarchy among right-handed neutrinos is much 
larger than the observed one among left-handed neutrinos.
Moreover, even under this assumption, we derive a precise
upper bound on the CP-asymmetry
and find it weaker than in previous literature~\cite{epsP}, leading to
a slightly higher neutrino mass bound.

Since this constraint is based on the doubtful assumption that
quasi-degenerate neutrinos
be produced by hierarchical right-handed
neutrinos, in section~\ref{sec:boundG}
we study what happens allowing quasi-degenerate
right-handed neutrinos.
As well known, the asymmetry can be resonantly enhanced~\cite{flanz,pil2};
with respect to the analysis of \cite{epsP}
we find other effects that relax the upper bound on the CP-asymmetry
so that our constraint on
neutrino masses is much weaker.
We discuss how the result depends on possible
reasons that naturally give rise to quasi-degenerate left-handed
neutrinos.

In section~\ref{T} we study alternative mechanisms of
leptogenesis and discuss the
corresponding constraints on neutrino masses.
Neutrino masses can be produced by tree-level exchange of
three different kinds of particles:
a) right-handed neutrinos \cite{seesawsinglet};
b) fermion $\hbox{SU}(2)_L$ triplets~\cite{tripletferm,ma};
c) one or more scalar triplets~\cite{scalartriplet}.
Special emphasis is put on the case b) which has never been
considered for leptogenesis. We show that it is efficient,
even if in this case the gauge interactions can keep triplets close to
thermal equilibrium.

Results are summarized in section~\ref{concl}.

%%%%%%%%%%%%%%%%%%%%%%%%%%%%%%%%%%%%%%%%%%%%%%%%%%%%%%%%%%%%%%%%%%%%%%%% 
%
\section{Hierarchical right-handed neutrinos}\label{H}
If neutrino masses  are produced
by the see-saw model described by the following Lagrangian
\beq\label{eq:Lseesaw}
\Lag = \Lag_{\rm SM} +\bar N_i i\partial\hspace{-1.3ex}/\, N_i +
(\lambdaN ^{ij} ~  N^i L^jH  +
\frac{M_N^{ij}}{2}   N_i N_j  +\hbox{h.c.})\,, \eeq
the most generic high energy parameters that give rise to
the desired neutrino masses $m_3>m_2>m_1\ge0$ and mixings $V$
can be written as~\cite{seesawparam}
\beq \label{eq:seesawParamCI}M_N =
\diag(M_1,M_2,M_3)\,,\qquad
\lambdaN  = \frac{1}{v} M_N^{1/2} \cdot R \cdot
\diag(m_{1},m_{2},m_{3})^{1/2}\cdot V^\dagger \,.  \eeq
One can always work in the mass eigenstate basis of
right-handed neutrinos, and choose
$M_3>M_2>M_1\ge0$.
$R$ is an arbitrary complex orthogonal matrix (i.e.\
$R^T\cdot R=1$),
that can be written in terms of 3 complex mixing angles as
\begin{equation}
R=\hbox{diag}\,(\pm1,\pm1,\pm1) \, R^{(23)}(z_{23}) R^{(13)}(z_{13})
R^{(12)}(z_{12}) \ ,  \label{eq:Rmatrix}
\end{equation}
where $R^{(ij)}$ is a rotation in the $(ij$) plane
with complex angles $z_{ij}$.
This parameterization explicitly shows
that the see-saw model has $12$ real and 6 complex parameters
beyond ones already present in the SM:
$6+3$ can be measured by low energy experiments
($m_1,m_2,m_3$ and the three complex mixing angles in $V$)
%,\theta_{12},\theta_{23},\theta_{13}$ and the CP-phases in $V$: $\phi,\alpha,\beta$) 
while the
remaining $6+3$ ($M_1, M_2, M_3$ and 
the three complex mixing angles in $R$:  $z_{12},z_{23},z_{13}$) cannot.

%%%%%%%%%%%%%%%%%%%%%%%%%%%%%%%%%%%%%%%%%%%%%%%%%%%%%%%%%%%%%%%%%%%%%%%
\subsection*{The CP asymmetry}
\begin{figure}[t]
$$\hspace{-4mm}\includegraphics[width=17.4cm]{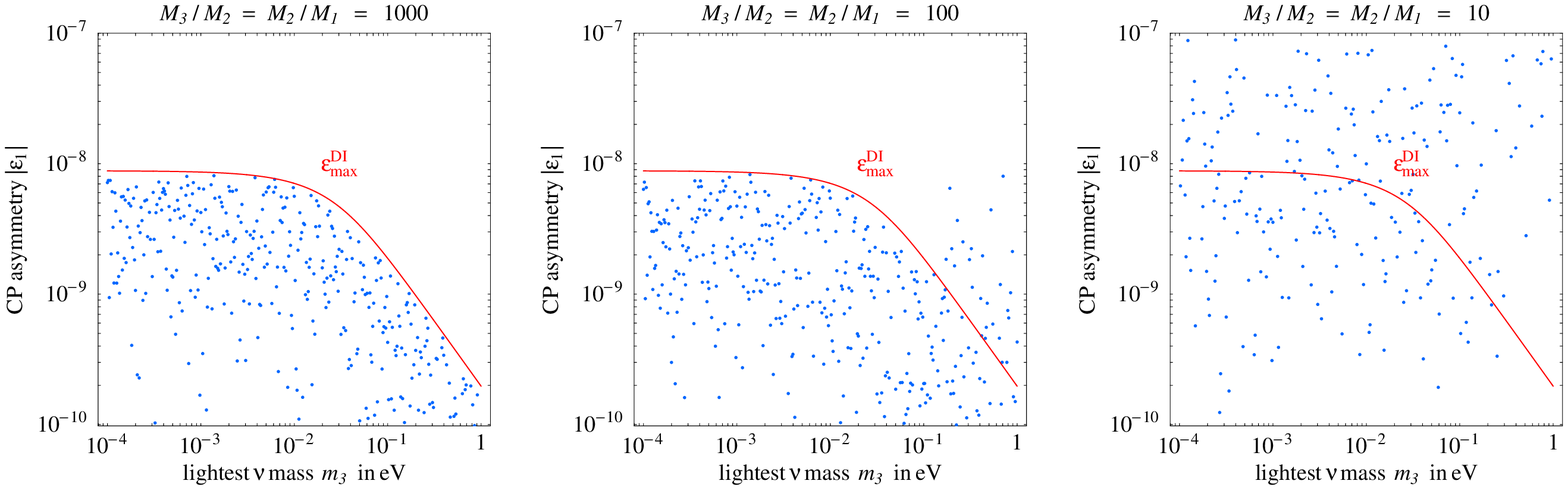}$$
\caption{\label{fig:EpsDI}\em
The Davidson-Ibarra bound on the CP asymmetry $\varepsilon_1$ was derived and
holds for $M_{2,3}/M_1 = \infty$.
The random samplings of the parameter space
(performed
assuming  different finite hierarchies among right-handed neutrinos)
confirms that $|\varepsilon_1|$ can be above the DI bound.
For all points $|\lambda_{ij}|<\sqrt{4\pi}$,
$\Delta m^2_{\rm atm} = 2~10^{-3}\eV^2$,
$M_1=10^8\GeV$.
As usual,
the density of points depends on 
arbitrary details of the sampling procedure
and carries no information about the 
likelihood of different regions.}
\end{figure}

One important ingredient that determines
the baryon asymmetry produced by thermal leptogenesis is the
CP asymmetry $\varepsilon_i$ in decays of right-handed neutrinos $N_i$.
Since we will later study the generic case where right-handed neutrinos  
can be
quasi-degenerate, it is useful to decompose the CP-asymmetry
in $N_1$ decays (and similarly for $N_{2,3}$ decays)
as the sum of a $V$ertex contribution %$\varepsilon_1^V$
and of a $S$elf-energy contribution (with the self energy contribution 
as given in \cite{pil2}) %$\varepsilon_1^S$
                              
\begin{equation}\label{eq:eps}
\varepsilon_1=-\sum_{j=2,3}\frac{3}{2}
  \frac{M_1 }{M_j }\frac{\Gamma_j }{M_j }
  I_j\frac{2 S_j + V_j}{3} \,,
\end{equation}
where
\begin{equation}\label{eq:IGamma}
I_j = \frac{ \hbox{Im}\,[ (\lambdaN  \lambdaN ^\dagger)_{1j}^2 ]}
{|\lambdaN \lambdaN ^\dagger |_{11} |\lambdaN \lambdaN ^\dagger |_{jj}}
 \, ,\qquad
\frac{\Gamma_j}{M_j} = \frac{|\lambdaN \lambdaN ^\dagger |_{jj}}{8\pi}
\equiv \frac{\tilde{m}_j M_j}{8\pi v^2}
\,,
\end{equation}
and where
\begin{equation}
S_j = \frac{M^2_j  \Delta M^2_{1j}}{(\Delta M^2_{1j})^2+M_1 ^2
   \Gamma_j ^2} \,, \qquad
V_j = 2 \frac{M^2_j }{M^2_1}
\bigg[ (1+\frac{M^2_j }{M^2_1})\log(1+\frac{M^2_1}{M^2_j })
- 1 \bigg]\,,
\end{equation}
are loop factors, with $\Delta M^2_{ij}=M^2_j-M^2_i$.
In the parameterization of eq.\eq{seesawParamCI},
the light neutrino mixing matrix $V$ does not affect $\varepsilon_1$
and CP-violation in leptogenesis arises from $R$.
While $V_j\le 1$, the factor $S_j$ is resonantly enhanced,
up to $S_j\sim M_j/\Gamma_j$, when $M_j-M_1\sim\Gamma_j$.
We normalized
the resonance factors $S_j$ and the vertex factors $V_j$ in
such a way that $S_{2,3}=1$
and $V_{2,3}=1$ (so that $(2S_j+V_j)/3=1$)
in the hierarchical limit $M_{2,3}/M_1\to\infty$.
In this limit, inserting the parameterization of eq.\eq{seesawParamCI}
into eq.\eq{IGamma} gives
  \begin{equation}\label{eq:epspreDI}
\varepsilon_1=-\sum_{j=2,3}
\frac{3}{16 \pi} \frac{M_1 }{M_j }
\frac{ \hbox{Im}\,[ (\lambdaN  \lambdaN ^\dagger)_{1j}^2 ]}
{|\lambdaN \lambdaN ^\dagger |_{11} }=-
  \frac{3}{16\pi}\frac{M_1}{v^2}
  \frac{\sum_i m_{i}^2 \hbox{Im}\, R_{1i}^2}{\sum_i
m_{i} |R_{1i}|^2} \,,
\end{equation}
which leads to the Davidson-Ibarra (DI) bound~\cite{leptogenesisBounds}:
\beq\label{eq:epsDI}  |\varepsilon_1| \le \varepsilon_{\rm max}^{\rm  
DI}=
  \frac{3}{16\pi}\frac{M_1}{v^2}(m_{3} - m_{1})\,, \eeq 
where $m_{3}$ ($m_{1}$) is the mass of the heaviest (lightest)  
neutrino.
The DI upper bound is plotted in fig.\fig{EpsDI}, together with
a random sampling that turns out to find points above it.
In fact, the DI bound is derived and holds for $M_{2,3}/M_1=\infty$
while in fig.\fig{EpsDI} we assumed a finite hierarchy.
One expects that for $M_{2,3}\gg M_1$ the DI bound remains  
approximatively valid,
up to small corrections of relative order $(M_1/M_{2,3})^2$.
We now explain why this is not the case.

In the infinitely hierarchical limit $\varepsilon_1$ is given by
a sum over right-handed neutrinos weighted by $1/M_j$ exactly like
neutrino masses:
the resulting simple expression has special properties,
e.g.~in this limit
CP violation in leptogenesis  disappears when light neutrinos are degenerate.
The extra terms suppressed by $(M_1/M_{2,3})^2$
do not share this property: e.g.\ they do not vanish when neutrinos
are degenerate.
Moreover, even with hierarchical neutrinos, these extra terms can be enhanced by
$\tilde{m}_{2,3}/m_{2,3}$.
%       $$\tilde{m}_i\equiv
%         |\lambdaN \lambdaN ^\dagger|_{ii} \frac{v^2}{M_i }\,.$$
One gets points above the DI bound
when the enhancements overcompensate the $(M_1/M_{2,3})^2$ suppression.

This observation can be relevant for leptogenesis, since
this enhancement of $\varepsilon_1$
can be achieved without
introducing significant wash-out factors
in the dynamics of thermal leptogenesis.
In fact, $\Delta L=2$ washout scatterings mediated by
off-shell $N_{1,2,3}$ exchange
are controlled
by neutrino masses and do not depend on $\tilde{m}_{2,3}$
(and scatterings mediated by on-shell $N_{2,3}$ are Boltzmann suppressed).
On the contrary $\Delta L=2$ scatterings mediated by on-shell $N_1$  
exchange
are controlled by $\tilde{m}_1$, and the efficiency of leptogenesis is  
maximal
for a relatively small value of $\tilde{m}_1\sim 
10^{-3}\eV$~\cite{epsP,thermal}.
Therefore we must show that the enhancement under discussion is
possible for small $\tilde{m}_1$.
We show this analytically and obtain simple estimates by
focussing on the simple limit $m_1=m_2=0$.
(Our final result remains valid in the more complicated
realistic case with small non-zero $m_1$ and $m_2$).
Inserting in the parameterization\eq{seesawParamCI}
$z_{23} = 0+i y_{23}$ and
$z_{13} = x_{13} + i y_{13}$ with $\cos2x_{13} = 1/\cosh 2y_{13}$ we get
\begin{equation}
\varepsilon_1 =- \frac{3}{16\pi}\frac{M_1 m_3}{v^2}(F_3 \cosh^2 y_{23}-
F_2 \sinh^2 y_{23}) \,,\qquad
\tilde{m}_1 = m_3 |\sin z_{13}|^2 = m_3 \frac{\cosh2y_{13}-\cos
   2x_{13}}{2} \,.
\end{equation}
Choosing a small $x_{13}$ allows to keep $\tilde{m}_1$ arbitrarily  
small without affecting $\varepsilon_1$.
In the fully hierarchical limit the loop functions $F_i = (2S_i+V_i)/3$
satisfy $F_2 = F_3=1$ and the last term in $\varepsilon_1$ simplifies  
to 1, giving $|\varepsilon_1| = \varepsilon_{\rm DI}^{\rm max}$.
For finite $M_1/M_{2,3}$ one instead has
$F_2\neq F_3$ and $\varepsilon_1$ can be enhanced by choosing a large  
$y_{23}$.
The maximal value is limited only by perturbativity of the largest
neutrino Yukawa coupling.
This violation of the DI bound
does not correspond to a local maximum of $\varepsilon_1$
and therefore was missed in analyses that tried to maximize $\varepsilon_1$ by imposing
$d\varepsilon_1/dz_{ij}=0$.
Imposing $\lambda_{33} = \cosh y_{23}
\sqrt{M_3 m_3}/{v}\circa{<}\sqrt{4\pi}$ gives
\begin{equation}\color{blus}\label{eq:stimaDI}
|\varepsilon_1| \circa{<} \max(\frac{M_1^3}{M_3M_2^2},
  \varepsilon_{\rm max}^{\rm DI}) \,.
\end{equation}
This estimate is confirmed by the random sampling
in figure\fig{EpsDI}, performed for $M_1=10^8\GeV$
which is probably the most interesting choice,
as the DI bound implies $M_1\circa{>}10^8\GeV$
(see e.g.\ \cite{thermal}).
Allowing only neutrino Yukawa couplings smaller than\footnote{In supersymmetric models where
sleptons acquire SUSY-breaking mass terms at high scale
the non-observation
of $\mu\to e\gamma$  
implies a
somewhat stronger bound,
$|\lambda^\dagger\lambda|_{e\mu}\circa{<}10^{-1\pm 1}$.}
$\sqrt{4\pi}$ and assuming $M_3/M_2 = M_2/M_1 = 10^n$ 
the new configuration under discussion allows
to reach $|\varepsilon_1|\circa{<} 10^{-4n}$.
The DI bound was derived and holds in the limit  $n\to\infty$.
For $n=3$ the DI bound is still an excellent approximation (fig.\fig{EpsDI}a).
For $n=2$ the DI bound starts failing only when neutrinos are quasi-degenerate (fig.\fig{EpsDI}b).
For $n=1$ the DI bound can be significantly evaded (fig.\fig{EpsDI}c).
This is possibly the most realistic case, as
solar and atmospheric oscillations indicate that there
is at most a mild hierarchy between left-handed neutrinos:  
$m_3/m_2\circa{<}6$.
Allowing right-handed neutrinos to have a similarly mild hierarchy,
$M_2/M_1 \sim 10$,
eq.\eq{stimaDI} shows that
successful leptogenesis with hierarchical $N_i$
is possible even for $N_1$ much lighter than $10^{8}\GeV$,
as possibly needed in supersymmetric models
in order to avoid overproduction of gravitinos~\cite{nucleo}.
Thermal leptogenesis can be successful even if right-handed neutrinos are light and hierarchical.
We can make a more quantitative statement by assuming a $10^{-3}$ efficiency 
(this is a reasonably conservative value, see e.g.~\cite{thermal}):
eq.\eq{stimaDI} then requires
$M_3 M_2^2/M_1^3\circa{<} 10^4 $ and consequently
allows some hierarchy among $M_{1,2,3}$.

\medskip

However, a CP-asymmetry above the DI bound is realized for
$\tilde{m}_{2,3}\gg m_{2,3}$ and therefore
involves apparently unlikely cancellations, as it needs that $N_2$ and  
$N_3$ each
gives a large contribution to neutrino masses,
but they cancel among each others.
This configuration seems justifiable in a natural way,
by e.g.\ building models where $N_2$ and $N_3$ form a quasi-Dirac couple
with quasi-chiral Yukawa couplings.
See also the appendix of~\cite{jarlskog}.

In the rest of this section we assume that $M_3M_2^2/M_1^3$ is large enough
for the DI upper bound to hold
and precisely compute, under all stated assumptions, the maximal
value of neutrino masses
compatible with thermal leptogenesis.
We find that previous bounds must be weakened
for a different reason, which has no relation with the above discussion.

%%%%%%%%%%%%%%%%%%%%%%%%%%%%%%%%%%%%%%%%%%%%%%%%%%%%%%%%%%%%%%%%%%%%%%%%
\subsection*{Bound on neutrino masses}

The DI bound becomes more stringent if neutrinos are quasi-degenerate  
since
in this case
$m_3-m_1\simeq \Delta m^2_{\rm atm}/2m_3$ in eq.\eq{epsDI} decreases.
Moreover in this case the efficiency factor $\eta$ is
smaller because larger neutrino masses need larger neutrino Yukawa  
couplings
and therefore implies larger wash-out scattering rates.
In fact the out-of equilibrium condition $\Gamma \circa{<} H(M_1)$
means $\tilde{m}_1\circa{<} 10^3 v^2/M_{\rm Pl}\sim 10^{-3}\eV$ where
$$\tilde{m}_1\equiv 
  |\lambdaN \lambdaN ^\dagger|_{11} \frac{v^2}{M_1 }=
8\pi \Gamma_1 \frac{v^2}{M_1 ^2}=
\sum_i m_{i} |R_{1i}|^2$$
is always larger than $m_1$.
The minimum value $\tilde{m}_1 = m_1$ implies $R_{12} = R_{13}=0$ and therefore
a vanishing CP-asymmetry (in the $M_{2,3}/M_1 = \infty$ limit).
Conversely, the DI bound is saturated for large values
of $\tilde{m}_1$, when $\eta$ is strongly suppressed.
As a result
 the maximal baryon asymmetry is reached for $\tilde{m}_1$
larger than $m_1\approx m_3$ but rather close to it~\cite{epsP}.

\begin{figure}[t]
$$
\includegraphics[width=8cm]{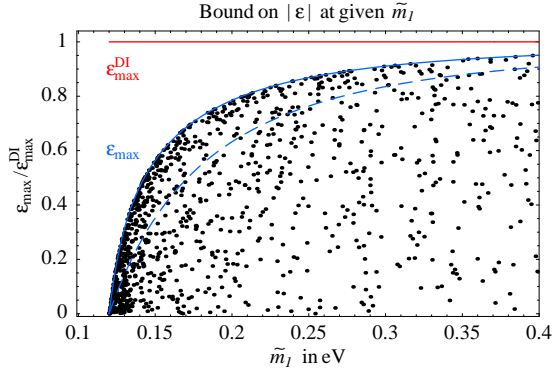}$$
\caption{\label{fig:EpsBound}\em
Maximal value of $\varepsilon_1$ assuming a large hierarchy at
fixed $\tilde{m}_1$.
The dashed line shows the attempt in~\cite{epsP}. The random sampling was
performed as an additional check that the analytical bound of eq.\eq{epsDeg} 
is correct.
}
\end{figure}

\begin{figure}[t]
$$\includegraphics[width=0.98\textwidth]{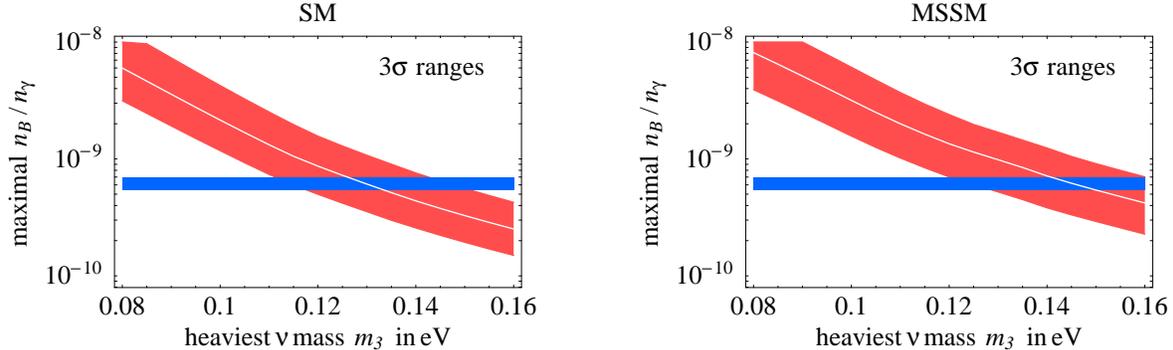} $$
\caption{\em\label{fig:m3}
{\bf Leptogenesis constraint on neutrino masses} assuming $M_{2,3}\gg  
M_1$.
The plot shows the measured baryon asymmetry (horizontal line)
compared with the maximal leptogenesis value as function of
the heaviest neutrino mass $m_3$, renormalized at low energy.
Error bars are at $3\sigma$.  }
\end{figure}

In order to compute the leptogenesis constraint
on neutrino masses, one must
compute the maximal value of $\varepsilon_1$ at given $\tilde{m}_1$,
and next maximize the baryon asymmetry $n_B/n_\gamma \approx 0.01  
\varepsilon_1\eta$
with respect to remaining free parameters,
essentially $\tilde{m}_1$ and $M_1$. To determine
the bound on $\varepsilon_1$ at fixed
$\tilde{m}_1$
we can neglect $\Delta m^2_{\rm sun}\ll \Delta m^2_{\rm atm}$,
so  that $m_1=m_2$ and we end up with a 2 neutrino case:
rotations in the (12) plane do not have physical effects.\footnote{One  
can explicitly verify that they do not affect $\varepsilon_1$.
From its explicit expression in eq.\eq{epspreDI}
the matrix $R_{12}$ in $\lambdaN $
commutes with $\hbox{\rm diag}\,(m_1,m_2,m_3)$ and cancels out with
$R_{12}^T$ coming from $\lambdaN ^T$.
This holds for the
numerator of eq.\eq{epspreDI}
because we assumed $M_{2,3}/M_1 = \infty$ and will
be no longer true when we will relax this assumption.
% To be more precise, the
The denominator has a dependence on $\hbox{Im} (z_{12})$, but
only as $c +c' \cosh \left[\hbox{Im}
(z_{12})\right]$ with $c$, $c'$ positive. The maximization of  
$\varepsilon_1$
naturally leads to $\hbox{Im} (z_{12})=0$,
hence the validity of what stated.}
We can write $R$ as
$$ R =R^{(13)}(z_{13}) = \pmatrix{\cos z_{13}
&0&\sin z_{13}\cr
0&1&0\cr -\sin z_{13} &0 &\cos z_{13}}\ ,$$
because $\varepsilon_1$ does not depend on $z_{23}$.
Here $z_{13} = x+ iy$ is a complex mixing angle.
The condition $\tilde{m}_1 = m_1 |R_{11}|^2 + m_3
|R_{13}|^2$ fixes $x$
as function of $y$:
$$\cos 2x = \frac{2\tilde{m}_1 - (m_1 + m_3)\cosh 2y}{m_3 -
m_1} \,,$$
allowing to write the CP asymmetry as
$$ |\varepsilon_1| = \frac{3}{16\pi}\frac{M_1}{v^2}
\frac{m_3^2- m_1^2}{\tilde{m}_1}
|\hbox{Im}R_{11}^2|=
\frac{3}{16\pi}\frac{M_1}{v^2}
\frac{m_3^2- m_1^2}{2\tilde{m}_1}
\sinh2y\sqrt{1 - \bigg(\frac{2\tilde{m}_1 -
(m_1+m_3)\cosh 2y}{m_3-m_1}\bigg)^2}\,.$$
Maximizing with respect to $y$ gives our bound:
\beq\label{eq:epsDeg}\color{blus}
  |\varepsilon_1| \le \varepsilon_{\rm max}=
\frac{ \varepsilon_{\rm DI}^{\rm max}}{2}
\sqrt{1-[(1-a)\tilde{m}_1/ (m_3-m_1)]^2}
\sqrt{(1+a)^2-[(m_3+m_1)/ \tilde{m}_1]^2 } \,, \label{eq:realepsbound}\eeq
where
$$a = 2  \hbox{Re} \bigg[\frac{m_1 m_3}{\tilde{m}_1^2}\bigg]^{1/3}
\bigg[-1 - i \sqrt{\frac{\left(m_1^2+m_3^2 +
\tilde{m}_1^2\right)^3}{27 m_1^2 m_3^2 \tilde{m}_1^2} -1}\bigg]^{1/3} >  
0 \,.
$$
In the hierarchical and  quasi-degenerate light neutrino 
limits it simplifies to
\beq\label{eq:epsDegApprox}
  |\varepsilon_1| \le \varepsilon_{\rm max}\simeq  \varepsilon_{\rm  
DI}^{\rm max}\times
  \left\{\begin{array}{ll}
1-m_1/\tilde{m}_1 & \hbox{if $m_1\ll m_3$}\\
\sqrt{1- m_1^2/\tilde{m}_1^2}&
\hbox{if $m_1\simeq m_3$}
\end{array}\right. .
\eeq
The continuous line in fig.\fig{EpsBound} (plotted assuming $m_1 =  
0.12\eV$)
is our upper bound.
The dashed line shows the result of~\cite{epsP},
that first attempted to compute $\varepsilon_{\rm max}$.
Their bound in the quasi-degenerate limit  reduces to
$ \varepsilon_{\rm DI}^{\rm max}(1-m_1^2/\tilde{m}_1^2)$
and would be stronger than our bound, but
does not hold as confirmed by the numerical scanning.\footnote{We  
explain analytically the reason of the disagreement.
Ref.~\cite{epsP} made unjustified assumptions on the elements of $R$.
In their notation variables $x_i$ and $y_i$ were used (with
$i=1,2,3$), together with the correspondence:
$
{R_{1i}^2 m_i}/{(\lambda \lambda^{\dagger} )_{11}}\equiv x_i+ i y_i
$.
The maximal $|\varepsilon_1|$ is reached for maximal $y_3$.
In order to find this maximum they assume $x_2=y_2=x_3=0$.
The first two assumptions are correct in the
limit $\Delta m^2_{\rm sun}\ll
\Delta m^2_{\rm atm}$ and
correspond to our $z_{12}=0$.
But the extra assumption $x_3=0$ (which would corresponds to
$\hbox{Re} R_{13}^2=0$) is incorrect, and doing so one does not
get the true maximum. In our numerical example with quasi-degenerate
neutrinos the maximum is reached
for $x_3\approx 0.4$.}
In the hierarchical limit $m_1\ll m_3$ the difference becomes less  
relevant.
In the MSSM the CP-asymmetry, and consequently its upper limit,
is 2 times larger than in the SM.
The MSSM result in fig.\fig{m3}b is obtained using the Boltzmann equations of~\cite{thermal}.

\bigskip

Combining our revised bound on the CP-asymmetry with the
revised computation of the efficiency of leptogenesis of~\cite{thermal}\footnote{In appendix~\ref{fla} we explain why, when computing such constraints,
the full network of Boltzmann equations can be approximated with a  
single equation
for the total $B-L$ asymmetry,
as tacitly assumed in previous analyses.}
we get
\begin{equation}
m_{3} < 0.15\eV \,\,\,\,\,\, (\hbox{at $3\sigma$})
\end{equation}
as illustrated in fig.\fig{m3}.
This constraint holds under the
doubtful assumption
that hierarchical right-handed neutrinos
give quasi-degenerate left-handed neutrinos.

%%%%%%%%%%%%%%%%%%%%%%%%%%%%%%%%%%%%%%%%%%%%%%%%%%%%%%%%%%%%%%%%%%%%%%%% 
%%%

%%%%%%%%%%%%%%%%%%%%%%%%%%%%%%%%%%%%%%%%%%%%%%%%%%%%%%%%%%%%%%%%%%%%%%%%
\section{Quasi-degenerate right-handed neutrinos}\label{sec:boundG}
Successful thermal leptogenesis implies
interesting restrictions on the masses of quasi-degenerate neutrinos
under the hypothesis of hierarchical right-handed neutrinos.
This is a crucial but doubtful assumption.
In fact, one expects that quasi-degenerate
neutrinos  be more naturally  produced by quasi-degenerate
right-handed neutrinos
(rather than by an interplay between heavier $N_{2,3}$ with bigger
Yukawa couplings and lighter $N_1$ with smaller couplings).
In absence of simple predictive models
one might  na\"{\i}vely expect that left-handed and
right-handed neutrinos show similar levels of degeneracy.

In this section we study how much the
constraint on neutrino masses gets relaxed
when we make this kind of `reasonable' assumptions.
We think this is an interesting but qualitative issue.
Therefore (unlike in the hierarchical case) we do not attempt
to derive a precise absolute constraint.
Our results should however be a qualitatively correct approximation to  
it.
A fully precise discussion is anyhow prevented by the fact that,
in a generic see-saw model, quasi-degenerate
neutrinos are not even stable under radiative corrections.
%(the bound is saturated when the smallest neutrino Yukawa coupling is  $\sim 0.3$).

We start studying
simple particular cases
and later show that they catch the new relevant effects
that we need to consider.

%%%%%%%%%%%%%%%%%%%%%%%%%%%%%%%%%%%%%%%%%%%%%%%%%%%%%%%%%%%%%%%%%%%%%%%% 
%%%%
\subsection*{The conservative case}
We first study what happens for  $M_3\gg M_2  \sim M_1 $.
Using eq.\eq{eps}, in this limit $\varepsilon_1$ (and similarly 
$\varepsilon_2$) can be approximated as
\begin{equation}
|\varepsilon_1|=\frac{M_1 }{M_2 } \frac{\Gamma_{2}}{M_2 }
S_2 |I_2| < \frac{1}{2} \frac{M_1 }{M_2 } |I_2| \simeq \frac{1}{2} |I_2| \,,
\label{eq:epsmaxdeg}
\end{equation}
where the inequality is obtained by taking the resonance
condition $M_2 -M_1 =\Gamma_2/2$ which maximizes the resonance
factor $S_2$ and gives $S_2 = M_2 /2\Gamma_2$.
It is useful to compare eq.\eq{epsmaxdeg} with
eq.\eq{epspreDI}.
The DI upper bound in eq.\eq{epsDI}
can be rewritten schematically as the product of two
suppression factors,
\begin{equation}
\frac{3}{2}
\frac{M_1 }{M_j }\frac{\Gamma_j }{M_j }\qquad\hbox{times}\qquad
\frac{m_3-m_1}{\tilde{m}_j}
\label{eq:supprfactors}
\end{equation}
present for different physical reasons.
  The first factor is related to the fact that heavy particles with
  small couplings
give small effects
  and it is just the result of a na\"{\i}ve estimation of $\varepsilon_1$
performed dropping the flavour indices:
\begin{equation}
\varepsilon_1 \sim \frac{3}{16 \pi}
\frac{M_1 }{M_j }
\frac{\lambda_j^2\lambda_1^2}{\lambda_1^2} \sim
\frac{3}{2} \frac{M_1 }{M_j } \frac{\Gamma_j }{M_j } \,.
\label{eq:naiveepsilon}
\end{equation}
The second factor comes from a flavour subtelty.
In the hierarchical limit, due to the orthogonality
of the $R$ matrix in eq.\eq{seesawParamCI}, the $\lambda_j^2$ in the
numerator of the asymmetry is not proportional to $\Gamma_j$ (or to
$\tilde{m}_j$) as in eq.\eq{naiveepsilon} but to
the $m_3-m_1$ mass difference.
This
results in the
extra  $(m_3-m_1)/\tilde{m}_j$ suppression of the asymmetry,
which is significant when neutrinos are quasi-degenerate.
For example for
$m_1 \sim 1$~eV this suppression is at least
of order $[\Delta m^2_{\rm atm}/(m_3+m_1)]/\tilde{m}_j\sim   
%2\cdot
10^{-3}$.

Eq.\eq{epsmaxdeg} shows that, when  also right-handed neutrinos are
quasi-degenerate, none of these two suppressions are there.
The first one can be completely compensated
by the resonance factor $S_2$ which is 1 in the hierarchical case
and $ M_2 /2\Gamma_2$ at the resonance.
The second suppression disappears
because $\varepsilon_1$ is no longer directly related to neutrino  
masses,
so that it no longer vanishes when neutrinos are degenerate.
More technically,  $I_2$ (unlike eq.\eq{epspreDI})
is not suppressed by an orthogonality relation coming from the $R$ matrix.

The net result is that the CP asymmetry can be of order unity  
independently
on the magnitude of the neutrino masses. Whether it is of order unity
is controlled by the size of $I_2$ in eq.\eq{epsmaxdeg}. As shown in
details in appendix~\ref{I2},
$I_2$ can be easily of order unity except if $\tilde{m}_i$ are very  
close to
their minimum values (where $\varepsilon_1$ vanishes).

A formula that correctly estimates the maximal CP asymmetry
not only in the quasi-degenerate case we are considering
but also in the hierarchical limit and in intermediate cases
is obtained by multiplying our bound  
$ \varepsilon_{\rm max} $ of eq.\eq{epsDeg}
valid in the hierarchical limit with an appropriate rescaling  factor:
\begin{equation}\label{eq:conservative}
|\varepsilon_1| \circa{<} \varepsilon_{\rm max} \frac{S_2 m_3-m_1}{m_3  
-m_1} \,.
\end{equation}
This bound reduces to eq.\eq{realepsbound} in the hierarchical limit  
$m_3\gg m_1$.
In the quasi-degenerate limit, $m_3\simeq m_1$, it reproduces  
eq.\eq{epsmaxdeg} up to an order-one factor.
This is the maximal CP-asymmetry possible
if a  quasi-degeneracy in neutrino masses, $m_1\approx m_2\approx  
m_3\approx\tilde{m}_1\approx\tilde{m}_2\approx \tilde m_3$
arises accidentally, as can happen if no flavour symmetry acts on  
neutrinos.
In such a case one expects that the three $\tilde{m}_i$ are equal
only up to order-one factors.
The resulting bound on $m_\nu$ is given by the solid line in fig.\fig{mdeg}a.
%       (the shaded area indicates the difference with respect to
%       a more precise approximation obtained including $\varepsilon_2$ and
%       using the results of the appendix). 
A numerical sampling performed including ${\cal O}(1)$ factors 
reveals that\eq{conservative} can be reached, 
and in fact the most conservative bound would be even somewhat weaker.
Already for a $\sim 10\%$ degeneracy between $N_2$ and $N_1$
successful leptogenesis can occur for neutrinos heavier than 1 eV.

%%%%%%%%%%%%%%%%%%%%%%%%%%%%%%%%%%%%%%%%%%%%%%%%%%%%%%%%%%%%%%%%%%%%%%%% 
%%%%
\subsection*{A simple special limit}

Next we consider the case where the 3 right-handed neutrinos are
quasi-degenerate. It is interesting to consider first the case
where two neutrinos are exactly degenerate and quasi-degenerate to a
third one, i.e.~$M_1  \simeq M_2 =M_{3}$ and assuming
  $\Gamma_2=\Gamma_{3}$. In such a special case $S_2=S_3$ and the
asymmetry reduces to:
\begin{equation}
|\varepsilon_1|= \frac{M_1 }{M_2 } \frac{\Gamma_2}{M_2 }
S_2 |I_2+I_3| < \frac{1}{8 \pi} \frac{M_1}{v^2} (m_3-m_1) S_2 \,.
\end{equation}
At the resonance this gives
\begin{equation}
|\varepsilon_1|
= \frac{1}{2} \frac{M_1 }{M_2 }
|I_2+I_3| \simeq \frac{1}{2} |I_2+I_3|
   < \frac{1}{2}
\frac{(m_3-m_1)}{\tilde{m}_2} \,.
\label{eq:epsmax3deg}
\end{equation}
As a result the asymmetry is smaller than in the previous 2 quasi
degenerate right-handed neutrino
case
by {\em one} power of the orthogonality factor that suppresses   
$I_2+I_3$. In other words
$\varepsilon_1$ is enhanced by the resonance factor but
still suppressed when neutrinos are quasi-degenerate, by the
$(m_{3}-m_{1})/\tilde{m}_2=
\Delta m^2_{\rm atm}/\tilde{m}_2(m_3+m_1)$ factor already discussed  
above.

\begin{figure}[t]
$$\includegraphics[width=7.5cm]{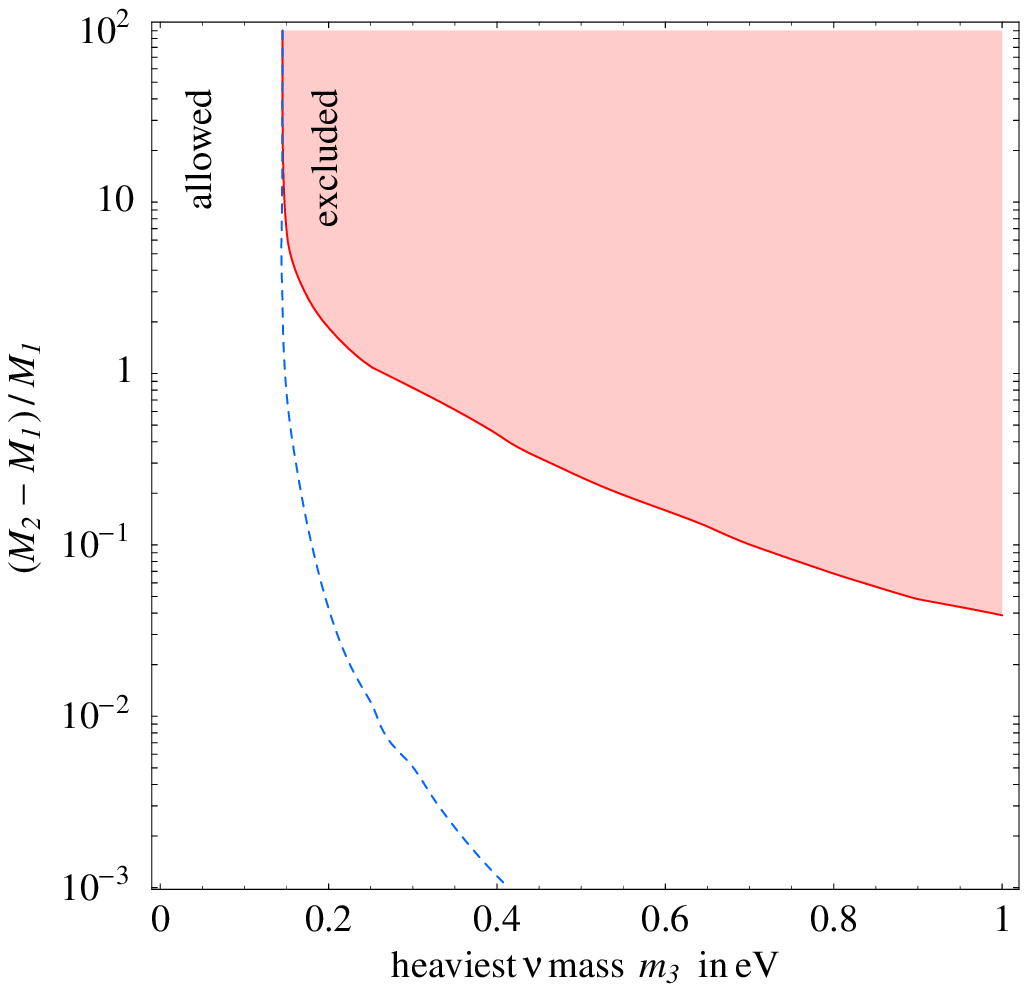}\hspace{8mm}
\includegraphics[width=7.5cm]{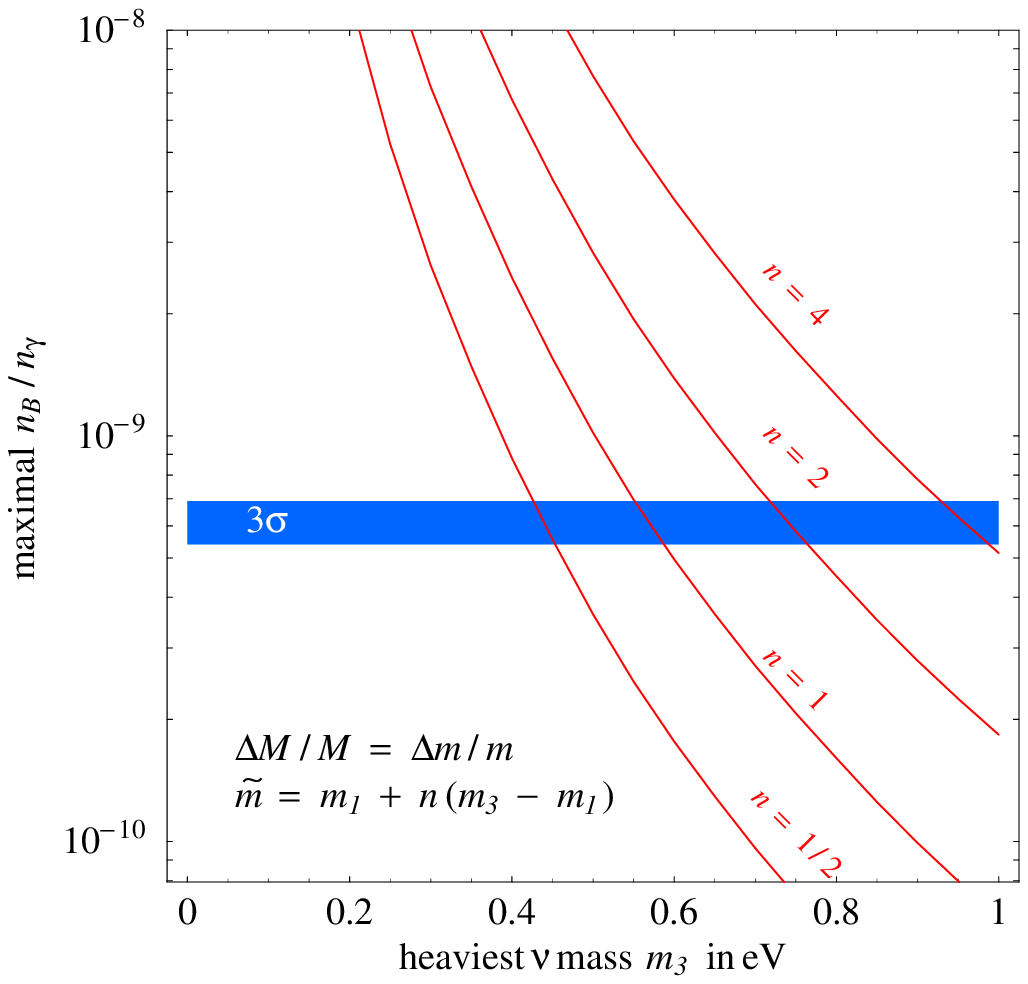}$$
\caption{\label{fig:mdeg}\em In fig.\fig{mdeg}a
we {\rm estimate} how much the
maximal value of neutrino mass compatible with thermal
leptogenesis increases
when right-handed neutrinos are allowed to be
quasi-degenerate.
The dashed line includes only the resonant enhancement of
CP-violation, eq.~\eq{epsmax3deg},
while the continuous line includes all effects.
A numerical sampling confirms that these constraints
can be saturated and even slightly exceeded.
Fig.\fig{mdeg}b holds 
in models where everything is as degenerate as neutrinos,
see eq.s~(\ref{sys:splits}).
The parameter $n$ quantifies how much $\tilde{m}_i$ are
assumed to be close to neutrino masses $m_i$.
As $n$ increases our assumptions get relaxed,
and therefore the constraint on $m_3$ becomes weaker.}
%so that $\tilde{m}$ must now be close to neutrino masses, 
%	considering
%	the natural possibility that the $\tilde{m}_i$ should be
%	quasi-degenerate with the neutrino masses (taking here for simplicity
%	$\tilde{m}_{1,2,3} \equiv \tilde{m}$ with
%	$\tilde{m}=m_1+n(m_3-m_1)$). 
\end{figure}

The constraint on neutrino masses
obtained in this special case
is shown by the dashed line in fig.\fig{mdeg}a
and is significantly stronger than
the one obtained in the previous case (continuous line).
This special case, which does not correspond to the most conservative
situation,
roughly corresponds to what is obtained in~\cite{epsP}.
In this reference, the hierarchical asymmetry has been enhanced 
by the resonance factor but was
still suppressed by the orthogonality factor (which
is $\sim 10^{-3}$ for $m_1 \sim 1$~eV).

%%%%%%%%%%%%%%%%%%%%%%%%%%%%%%%%%%%%%%%%%%%%%%%%%%%%%%%%%%%%%%%%%%%%%%%% 
%%
\subsection*{The most realistic case}
Since the leptogenesis  constraint on neutrino masses is relevant
  only for a quasi-degenerate spectrum of light neutrinos, one can
  wonder which is the most natural right-handed neutrino mass
  spectrum that produces quasi-degenerate neutrinos.
 Presumably the answer is: three  quasi-degenerate right-handed neutrinos.
 In fact, other spectra (e.g.\ hierarchical right-handed neutrinos)
can give rise to quasi-degenerate neutrinos
 only in presence of an appropriate precise correlation
 between the Yukawa couplings and the right-handed neutrino masses.
 It is difficult to find a theoretical reason that can justify 
 this kind of correlation among different objects.

With three 
quasi-degenerate right-handed neutrinos, no qualitatively new effect appears with respect to the
`two neutrino' quasi-degenerate case of eq.\eq{epsmaxdeg} we discussed above.
An orthogonality suppression similar to the one of the special case  
above
is generically not present.
As a consequence the constraint on neutrino masses
is again well estimated by the continuous line in fig.\fig{mdeg}a.

There exists one specific pattern,
which is probably the most realistic one, which leads
to more stringent bounds.
In fact, if neutrinos were quasi-degenerate,
the degeneracy would presumably not be accidental but due to
some reason:
a broken SO(3) flavour symmetry is probably the simplest possibility.
One expects that in such framework all quantities, and not only
neutrino masses, are close to the ideal limit
where three degenerate right-handed neutrinos
give equal masses (with equal CP phases)
to three orthogonal combinations of left-handed  
neutrinos.
Therefore one expects something like $\tilde{m}_i - \tilde{m}_j\approx m_i - m_j$ and
\begin{eqnsystem}{sys:splits}
&&\frac{M_2-M_1}{M_1}\sim
\frac{\tilde{m}_2-\tilde{m}_1}{\tilde{m_1}}\sim
\frac{m_2-m_1}{m_1} \approx \frac{\Delta m^2_{\rm sun}}{2 m_1^2}\approx
0.5~10^{-4} \bigg(\frac{\eV}{m_{2}}\bigg)^2, \label{eq:splitsol}\\
&&\frac{M_3-M_2}{M_2}\sim \frac{m_3-m_2}{m_2} \sim
\frac{\tilde{m}_3-\tilde{m}_2}{\tilde{m_2}}
\approx \frac{\Delta m^2_{\rm atm}}{2 m_3^2}\approx
10^{-3} \bigg(\frac{\eV}{m_{2}}\bigg)^2. \label{eq:splitatm}
\end{eqnsystem}
Right-handed neutrinos can be more degenerate than in the above  
estimates
if only the neutrino Yukawa couplings
deviate from the symmetric limit, and can be less degenerate only
if there are accidental cancellations between non-universal
Yukawa couplings and non-degenerate $M_{1,2,3}$
in the see-saw prediction for neutrino masses.
%  ...if only .... only if... e'giusto cosi'. Non'e'una svista.
Within the parameterization of eq.\eq{seesawParamCI}, our reasonable  
assumption means
$|y_{ij}|\circa{<} \Delta m^2/m^2$.

To calculate the bound on the neutrino mass considering the realistic
example of eq.s~(\ref{sys:splits}), it is an excellent
approximation to take in eq.\eq{eps} (and similar equation for $N_{2,3}$) all
$M_i$ equal to the same value $M$ everywhere except in the resonance
$S$ factors, and to take all decay widths equal to the same value
$\Gamma$ everywhere except in the Yukawa coupling $I$ factors. It 
is also an excellent  
approximation to
calculate the
wash-out effects in the symmetric limits with
$\tilde{m}_1=\tilde{m}_2=\tilde{m}_3\equiv\tilde{m}$.
Making these approximations
(and assuming $M_{1,2,3}\gg10^{10}\GeV$, so that 
interaction rates induced by
charged lepton Yukawa couplings can be neglected)
the complicated set of Boltzmann equations for
the lepton asymmetries generated by $N_{1,2,3}$ decays
splits into three independent  Boltzmann equations.
The net result is that the efficiency is the same as
in `one flavour' approximation, with the CP-asymmetry now given by
$\varepsilon=\varepsilon_1+\varepsilon_2+\varepsilon_3$
(which can be rewritten in a rephasing-invariant way as a trace
of an appropriate matrix function of $\lambda$ and $M_N$, see~\cite{jarlskog}).
In this case in the limit $M_1=M_2$ the asymmetry reduces to the one
of the special case above which is orthogonality suppressed by factors 
$(m_3-m_1)/\tilde{m}_i$. Therefore the only terms which are not
orthogonality suppressed and which are dominant are the terms
involving
the $M_2^2-M_1^2$ splitting. Neglecting higher order terms in the splitting
parameters $(M_3-M_1)/M_1$, $(M_2-M_1)/M_1$ and $(M_2-M_1)/(M_3-M_1)$
these terms are:
\begin{equation}
%\varepsilon \simeq 2 \frac{\Gamma}{M}- S_2 I_2 - 2 \frac{\Gamma}{M} S_3
%S_3 \frac{M^2_2-M^2_1}{M_3^2-M_1^2} 
%\frac{(M_3^2-M_1^2)^2- M^2 \Gamma^2}{(M_3^2-M_1^2)^2+M^2 \Gamma^2}  
\varepsilon  \simeq - 2 \frac{\Gamma}{M} S_2 I_2 + 2
\frac{\Gamma}{M} S_3
I_3 \frac{M_2-M_1}{M_3-M_1}\frac{(M_3-M_1)^2-
\Gamma^2/4}{(M_3-M_1)^2+ \Gamma^2/4}  \,.
\label{eq:epsreal}
\end{equation}
The first term is the asymmetry of the 2 quasi
degenerate case of eq.\eq{epsmaxdeg} (from both $\varepsilon_1$ and
$\varepsilon_2$) in which $N_3$ has a negligible effect.
The second term comes from the difference between the $N_1$-$N_3$
and $N_2$-$N_3$ mass splittings in the diagram involving these
right-handed neutrinos. It is easy to check that although this term
can have large effects for large $M$,\footnote{At large $M$
  (e.g.~above $10^{13\div14}$~GeV), unlike for smaller $M$, the sum
  of both terms is suppressed by orthogonality $(m_3-m_1)/\tilde{m}_i$ 
  factors, and the corresponding neutrino mass bound is therefore suppressed.}
% (e.g. above $10^{13}$~GeV), 
it 
has a completely negligible effect
on the bound which is obtained for smaller $M$.
The asymmetry relevant for the determination of the bound
reduces therefore to the 2 quasi-degenerate
case above of eq.\eq{epsmaxdeg} summed on both $\varepsilon_1$ and
$\varepsilon_2$.
In addition to the fact that
it is not
suppressed by any orthogonality relation factor, it turns out to be
little suppressed by the resonance factor $S_2$.
For values of
$m_1$ around eV and with $\tilde{m}_j \sim m_i$ the 
factor $S_2$ is naturally at
the resonance or close to it.
From eq.\eq{splitsol}, the resonance
condition $\Gamma \sim 2(M_2-M_1)$ gives:
\begin{equation}\label{eq:Mdeg}
M \simeq 10^{11} \,\, \hbox{GeV} \Big( \frac{\eV}{m_i} \Big)^3 \,,
\end{equation}
The only large suppression can come from the factor
$I_2$ for values of the $\tilde{m}_i$ close to $m_1$.
From the bound on $I_2$ given in appendix~\ref{I2}, this suppression goes  
like:
\begin{equation}
|\varepsilon_{\rm max}|=I_2^{\rm max}
\simeq (1-m_1/\tilde{m})^{3/2} \,,
\label{eq:I2suppress}
\end{equation}
where for simplicity in the last equality we have taken all
$\tilde{m}_i$
equal to the same value $\tilde{m}$.
Using this bound, in fig.\fig{mdeg}b we give the
baryon asymmetry we obtain as a function of $m_3$ for values of
$\tilde{m}=m_1+n(m_3-m_1)$ with $n=\{1/2,1,2,4\}$. Taking  
$\tilde{m}=m_3$ ($n=1$),
as the generic example for the case that the $\tilde{m}$ would be
precisely of order the neutrino masses, gives the constraint
\begin{equation}
m_3 < 0.6 \,\, \hbox{eV}  \label{eq:m3reali}
\end{equation}
which is stronger than in the conservative case %(continuous line of fig.\fig{mdeg}a)
because we are now assuming smaller $\tilde{m}$, close to neutrino masses.
%while $n=1/2$ gives $m_3<0.45$~eV.
Taking larger values of $\tilde{m}$ leads rapidly to larger bounds. For
example taking $n=4$ (which starts be fine-tuned)
gives $m_3<1$~eV.
The dependence of the bound on $M_1$ is quite sensitive to the exact
value of the splitting we take for the right-handed neutrinos because  
this
determines the position of the resonance. For 
eq.s~(\ref{sys:splits}) the
bound is obtained for values close to where the $N_1/N_2$
resonance occurs, i.e.~around  $M\sim10^{11}$~GeV.
Without a predictive flavour model which would show how
the correlations between the seesaw parameters
at the origin of the degenerate spectrum occur, these bounds obtained
from eq.s~(\ref{sys:splits}) are only indicative of what  
happens and
in order to have a safe bound we must consider the conservative case
of fig.\fig{mdeg}a
 (where $n$ was left as a free parameter in order to maximize the asymmetry, 
so that $\tilde{m}_j$ can differ from $m_i$ by factors of order one).
Even in a very constrained situation the
neutrino masses
can be as large as 0.6~eV, eq.\eq{m3reali}.\footnote{Stronger constraints will arise if supersymmetry exists
and if right-handed neutrinos lighter than in eq.\eq{Mdeg}
will be needed to avoid gravitino overproduction~\cite{nucleo}.
Furthermore, making extra `reasonable' assumptions about the flavour
structure of Yukawa couplings one gets
a smaller CP-asymmetry (suppressed by 2 powers of the quasi-degeneracy factor,
rather than by 3/2 powers) and consequently a slightly stronger constraint on neutrino masses.}

%%%%%%%%%%%%%%%%%%%%%%%%%%%%%%%%%%%%%%%%%%%%%%%%%%%%%%%%%%%%%%%%%%%%%%%% 
%%%%%
\section{Leptogenesis in alternative minimal models of neutrino  
masses}\label{T}
Generic neutrino masses can be mediated by tree-level exchange of:
\begin{itemize}
\item[a)] At least three fermion singlets
(`right-handed neutrinos'), described by the see-saw Lagrangian of  
eq.\eq{Lseesaw}.

\item[b)] At least three fermion SU(2)$_L$ triplets with zero hypercharges:
the Lagrangian keeps the same structure as in the singlet case,
but with different contractions of the SU$(2)_L$ indices that we  
explicitly show:
\beq\label{eq:Lseesaw3}
\Lag = \Lag_{\rm SM} +\bar N_i iD\hspace{-1.5ex}/\, N_i +
(\lambdaN ^{ij} ~  \tau^a_{\alpha\beta} N^a_i  
L_j^\alpha H^\beta  + \frac{M_N^{ij}}{2}   N_i^a N_j^a+\hbox{h.c.}).  \eeq
The index $a$ runs over $\{1,2,3\}$,
$\alpha,\beta$ over $\{1,2\}$ and
$\tau^{a}$ are the Pauli matrices.

\item[c)] One scalar (`Higgs') triplet $T$ with
appropriate hypercharge, such that
the most generic renormalizable Lagrangian is
\beq\label{eq:tripletS}\Lag = \Lag_{\rm SM} + |D_\mu T|^2- M_T^2 |T|^2+
({\lambda}^{ij}_T L^iL^j T  + M\, HH T^*+\hbox{h.c.} ).\eeq

\end{itemize}
Neutrino masses can be also mediated by combinations of the above possibilities,
among which it is interesting to consider:
\begin{itemize}
\item[d)] Two or more scalar triplets $T$ with similar interactions.
\item[e)] One scalar triplet and fermion singlets.
\end{itemize}
Model c) has the minimal number of beyond-the-SM parameters
(8+3, while a) and b) have both 12+6 extra parameters)
but does not lead to a large enough lepton asymmetry.\footnote{One  
expects
that a CP asymmetry in the total triplet
decay rate, $\Gamma(T\to LL)\neq \Gamma(T^*\to\bar L\bar L)$
arises at two or more loops.
Taking into account
how $\lambda_T$ and the Yukawa couplings of charged leptons
break  U(3)$_L\otimes{\rm U}(3)_E$ flavour rotations
and proceeding along the lines of~\cite{dipoli}
we find that a non zero CP asymmetry needs four powers of
$\lambda_\tau$ and two powers of $\lambda_\mu$, and is therefore
too small (unless enhanced by IR effects, which might
give only a mild logarithmic GIM-like suppression).}
Adding to c) other scalar triplets as in d) or fermion singlets as in e) 
allows successful leptogenesis\footnote{Leptogenesis in case d) has  
been studied
in~\cite{masar,scaltriplepto}. We do not consider further this
possibility here.} at the price of introducing more unknown parameters
than in a), b) and c).
Theoretically, case a) is
preferred because singlets
nicely marry with grand unification
(which is maybe the most promising speculation that we have today).
The combination e) can also find theoretical support because
3 singlets and a Higgs triplet are naturally present in
renormalizable SO(10) models (as well as the in underlying left-right  
models)
and can play an important r\^ole in leptogenesis~\cite{sod,hs,lazdent}. This
possibility does not lead to relevant constraints on neutrino mass.
%\footnote{In renormalizable
%$SO(10)$ or left-right models, both a) and c) can play a role for
%leptogenesis \cite{hs}. Fermion triplets can be present if there is
%for example $ 24$ multiplets of SU(5)~\cite{ma}.}.
Cases b) and d) seem to be less natural within a grand
unified scheme.
E.g.\ fermion triplets with the needed Yukawa  
couplings
can arise from adjoints of SU(5).
However we believe that it is worth to study case b) because it
is the only possibility which, with as few parameters as the singlet
model,
can lead to successful
leptogenesis.
This is what we show in the following.

\begin{figure}[t]
$$
\includegraphics[width=16cm]{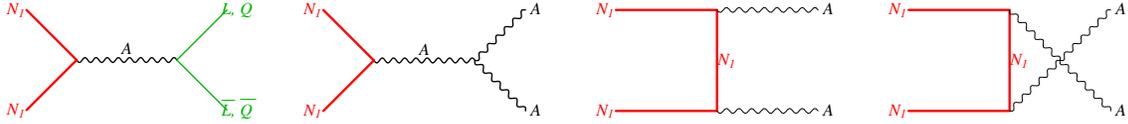}$$
\caption{\label{fig:FeynT}\em
{\bf Fermion triplet leptogenesis}. 
Feynman diagrams that give the new interaction rate $\gamma_A$.}
\end{figure}

\begin{figure}[t]
$$
\includegraphics[height=8cm]{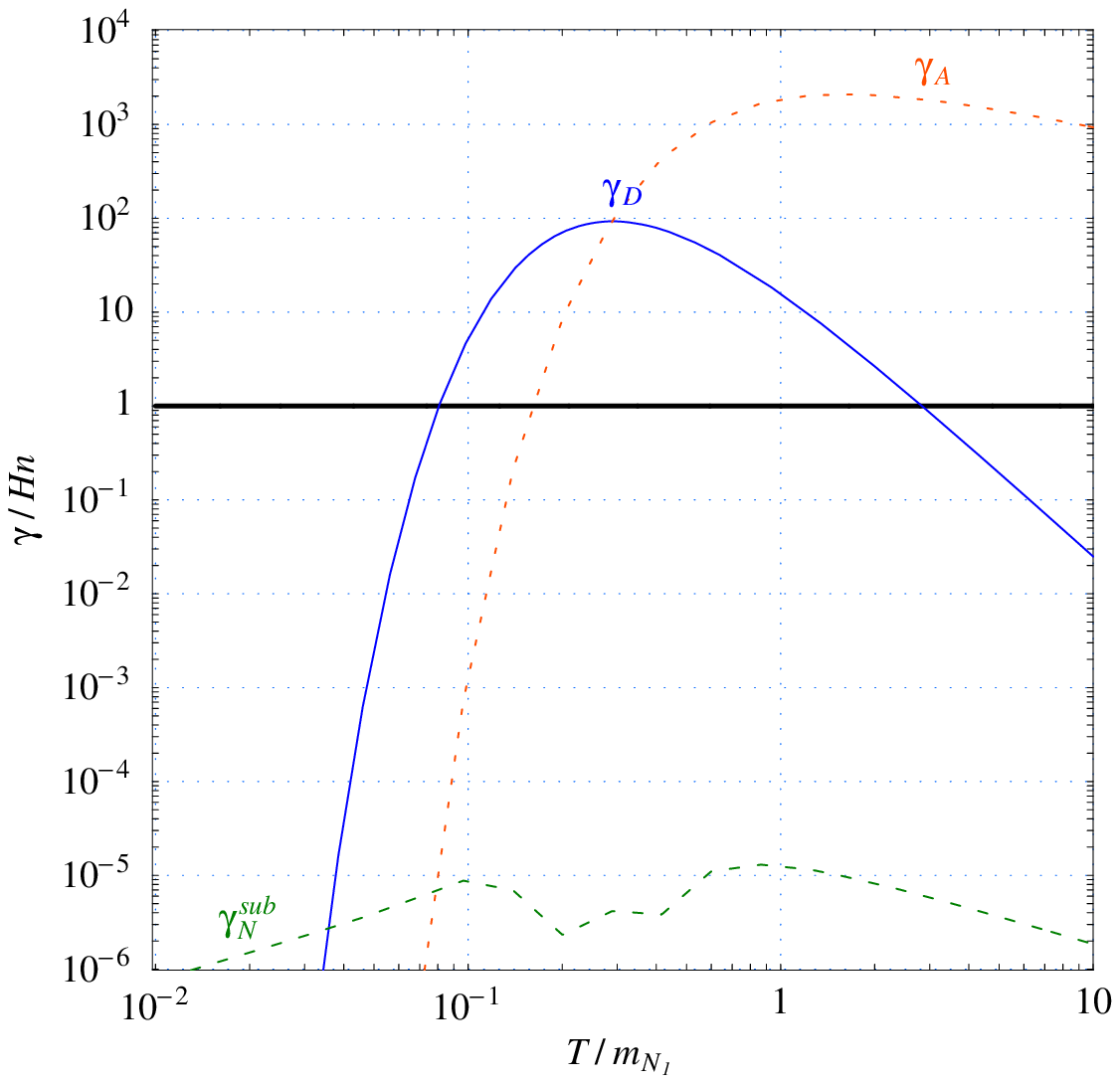}~~~
\includegraphics[height =8cm]{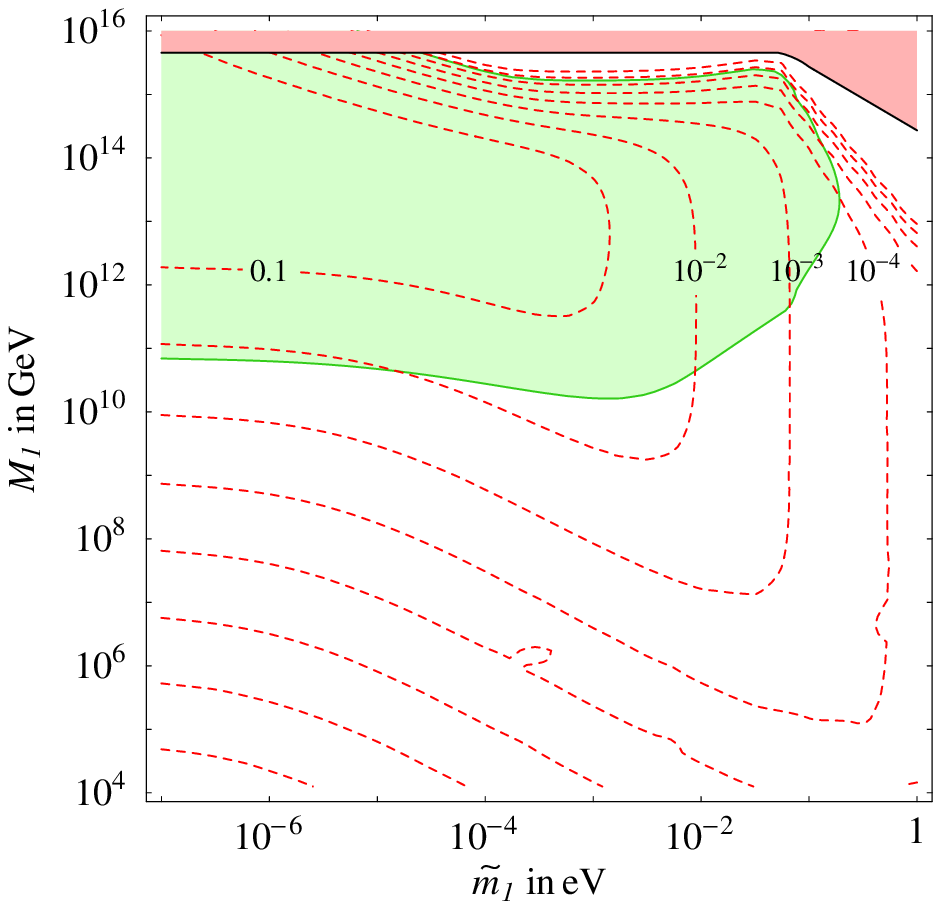}$$
\caption{\label{fig:etaT}\em
{\bf Fermion triplet leptogenesis}. 
Fig.\fig{etaT}a: Interaction rates $|\gamma/Hn_\gamma|$
for $M_1 = 10^{10}\GeV$ and $\tilde{m}_1 = 0.06\eV$.
Fig.\fig{etaT}b: contour-levels of the efficiency $\eta$.
Successful leptogenesis with (infinitely)
hierarchical triplets is possible inside the green area.
}
\end{figure}

Neutrino singlets trivially allow thermal leptogenesis:
not having gauge interactions they
easily satisfy the out-of-equilibrium Sacharov condition for  
baryogenesis.
Fermion triplets (as scalar triplets) have gauge interactions so that
it is more difficult to have a  non thermal abundancy.
We now study thermal leptogenesis in decays of charged particles,
finding that gauge interactions, rather than preventing baryogenesis,
make it more predictive.
The point is that gauge interactions involve two particles (see fig.\fig{FeynT})
and are therefore doubly Boltzmann suppressed at temperatures below  
their mass
(fig.\fig{etaT}a shows an explicit example), so that
they cannot wash-out the lepton asymmetry in an efficient
way.\footnote{A similar result was found for
scalar triplets in~\cite{scaltriplepto}.}
On the contrary gauge interactions are efficient at higher temperatures
and thermalize the initial abundancy, so that the final baryon  
asymmetry
almost never depends on it
(unlike what happens in the singlet case).

With fermion triplets neutrino masses are
still given by the usual see-saw formula,
$m = -v^2 \lambda^T \cdot M_N^{-1} \cdot \lambda$,
without changing any ${\cal O}(1)$ factor.
Using the same notations as in the singlet case (see eq.\eq{eps} for a more precise discussion)
the CP asymmetry is now given by
\begin{equation}\label{eq:epsT}
\varepsilon_1=\sum_{j=2,3}\frac{3}{2}
  \frac{M_1 }{M_j }\frac{\Gamma_j }{M_j }
  I_j\frac{V_j-2 S_j }{3} \,,
\end{equation}
and is therefore 3 times smaller in the hierarchical limit.
The final amount of baryon asymmetry is given by
the CP-asymmetry times the efficiency factor $\eta$ times a
numerical coefficient which is  3 times bigger than in the singlet case  
because now $N_1$ has three components:
\begin{equation}
\label{uusm}
\frac{n_{B}}{n_\gamma} =-0.029 \varepsilon_{1} \eta \,.
\end{equation}
The $N_1$ decay width is given by the same expression as in the singlet  
case,
so that the thermally averaged decay rate $\gamma_D$ becomes 3 times  
bigger
(again because $N_1$ now has 3 components).
The on-shell part of $\Delta L=2$ scattering rates, equal
to $\gamma_D/4$~\cite{thermal},
becomes therefore also 3 times bigger, and the off-shell part is  
affected in a different way.
We find:
\begin{eqnsystem}{sys:scatt}
\hat\sigma_{Ns} (LH\to\bar L H^*)&=& \frac{(\lambda \lambda^\dagger)_{11}^2}{4\pi}\bigg[\nonumber
2+x D_s^{\rm 2sub}+(2-3x\xi )\hbox{Re}D_s+3\xi (x\xi -2)-\\
&&-\frac{2\ln(1+x)}{x}(1-(\hbox{Re}D_s-3\xi )(1+x))\bigg]\\
\hat\sigma_{Nt} (LL\to H^* H^*)&=& \frac{(\lambda \lambda^\dagger)_{11}^2}{2\pi}\bigg[\frac{3x}{2}(\xi ^2+\frac{2}{1+x})+
(3\xi -\frac{3}{2+x})\ln(1+x)\bigg]\\
\riga{where $x=s/M_1^2$.
A `natural' value of the parameter $\xi$ is
$\xi = m_3/\tilde{m}_1$. 
It is defined as follows: the amplitude of $N_{2,3}$-mediated scatterings is 
written as $\xi$ times the value computed assuming that $N_{2,3}$ give the same neutrino masses as $N_1$.
In order to deal with this issue in a more precise way
one should know the flavour structure of $N_1$ couplings
and solve Boltzmann equations for the asymmetries in the various flavours.
Our simplified approach is justified by the fact that 
$\xi$ has a minor impact in most of the `reasonable' parameter space.
The reduced cross sections $\hat\sigma$ are related to the corresponding interaction rates
as summarized in~\cite{thermal}, that also explains how to perform
a proper subtraction of the $s$-channel propagator $D_s$.

\smallskip

We computed gauge scatterings $\hat{\sigma}_A$ (see fig.\fig{FeynT})
summing over the 12 SM fermionic doublets $D=\{L_{1,2,3},Q_{1,2,3}\}$
and neglecting scatterings into Higgs doublets 
(since they are not enhanced by a large
number of final states and since the threshold behavior at $s\simeq M_1^2$ is the same).
At $s\gg M_1^2$ the $NN\to AA$ cross section is enhanced by IR effects.
We find:
}\\
\hat\sigma_A(N_1N_1\to D\bar{D},AA)&=& \frac{6 g_2^4}{\pi }(1+\frac{2}{x})r +
%\frac{g_2^4}{4\pi} \bigg[-r(31 +\frac{134}{x}) + 24(1+\frac{4}{x}-\frac{4}{x^2})\ln\frac{1+r}{1-r}\bigg]
\frac{2g_2^4}{\pi} \bigg[-r(4 +\frac{17}{x}) + 3(1+\frac{4}{x}-\frac{4}{x^2})\ln\frac{1+r}{1-r}\bigg]
\end{eqnsystem}
where $r= \sqrt{1-4/x}$.
Symmetry factors for initial and final state particles are included in the reduced cross sections. 
For simplicity we neglected $\Delta L=1$ scatterings, three body decays,
one loop and thermal corrections.

The Boltzmann equation involve the new gauge term, $\gamma_A$:
   \begin{eqnsystem}{sys:Boltz}
sHz \frac{dY_{N_1}}{dz} &=&
  -\bigg(\frac{Y_{N_1}}{Y_{N_1}^{\rm eq}}-1\bigg)\gamma_D
  -\bigg(\frac{Y_{N_1}^2}{Y_{N_1}^{2\rm eq}}-1\bigg)\gamma_A \,, \\
sHz \frac{dY_{{\cal B} - {\cal L}}}{dz} &=&
-\gamma_D \varepsilon_{N_1} \bigg(\frac{Y_{N_1}}{Y_{N_1}^{\rm eq}}-1\bigg)  
-\frac{Y_{{\cal B} - {\cal L}}}{Y_{L}^{\rm eq}}\bigg(\frac{\gamma_D}{2}+
2\gamma_{N}^{\rm sub}\bigg) \,,
  \label{eq:BoltzB-L}
\end{eqnsystem}
  where $s$ is the entropy of SM  
particles,
$H$ is the Hubble constant at temperature $T$.

The efficiency factor $\eta(\tilde{m}_1, M_1,\xi)$,
computed solving numerically the Boltzmann equations,
is shown in fig.\fig{etaT}b, assuming
the reasonable value of $\xi$
(other reasonable values would give minor differences).
The result is qualitatively different from the analogous
result in the singlet case, shown in fig.~8 of~\cite{thermal}:
\begin{itemize}
\item At $\tilde{m}_1\ll 10^{-3}\eV$ only  gauge interactions drive the  
$N_1$ abundancy close
to thermal equilibrium (unless $M_1\circa{>} 10^{15}\GeV$).
A `relic' fraction of $N_1$ survives to gauge annihilations
and later decays generating the baryon asymmetry with efficiency
$\eta \approx M_1/10^{13}\GeV$
($\eta$ is larger if $\tilde{m}_1\sim 10^{-3}\eV$ because
some $N_1$ decay during the annihilation stage).
Gauge interactions give a stronger suppression at smaller $M_1$, because at low temperatures
the expansion of the universe is slower, $H\sim T^2/M_{\rm Pl}$.
This can be contrasted to what happens in the $\nu_R$ case:
since it has no  gauge interactions the efficiency $\eta$ only has a  
minor dependence on $M_1$.
\item At  $\tilde{m}_1\gg 10^{-3}\eV$ neutrino Yukawa interactions  
drive the $N_1$ abundancy close
to thermal equilibrium.
As a consequence there are only ${\cal O}(1)$ differences between
singlet and triplet leptogenesis. 
\end{itemize}
Assuming a sufficiently huge hierarchy, $M_1\gg M_{2,3}$ we can
derive a region where thermal leptogenesis can be successful
using our maximal CP-asymmetry
(which is $1/3$ of what obtained in eq.\eq{epsDeg} in the singlet case).
The lower bound on the $N_1$ mass and the upper bound
on neutrino masses are
at $3\sigma$
\begin{equation}\label{eq:Tripletbound}
M_1\circa{>}1.5~10^{10}\GeV\qquad m_3 < 0.12\eV\,.
\end{equation}
These bounds are slightly stronger than in the right-handed neutrino
case\footnote{Within one order of magnitude, our bound
on $M_1$ is in agreement with the estimated
   bound in~\cite{scaltriplepto} for scalar triplets (see
also~\cite{hs}).},
and are subject to all the caveats discussed in that case.
In particular quasi-degenerate $N_i$ allow leptogenesis
at the TeV-scale: fig.\fig{etaT}b shows that, despite gauge interactions, the efficiency
remains large enough.
This case is testable at collider,
where $N_i$ triplets can be produced and detected
(while $N_i$ singlets cannot, because have too low cross sections).

Only ${\cal O}(1)$ factors are modified if
supersymmetry is introduced in the usual minimal way.

%%%%%%%%%%%%%%%%%%%%%%%%%%%%%%%%%%%%%%%%%%%%%%%%%%%%%%%%%%%%%%%%%%%%%%%
\section{Conclusions}\label{concl}
Atmospheric oscillations suggest that the heaviest neutrino mass $m_3$
is larger than about $0.05\eV$.
Various techniques could reach the necessary sensitivity
and presently give the following 95\% C.L.\ bounds:
%time delay from 1987A supernova implies $m_3\circa{<} 10\eV$,
$m_3 < 2.2\eV$ from $\beta$-decay~\cite{beta},
$m_3<1.0 h~\eV$ from neutrino-less double-beta decay~\cite{0n2b}
(assuming Majorana masses; $h\approx 1$ renormalizes the uncertain  
nuclear matrix element),
$m_3 < 0.23\eV$ from cosmology~\cite{WMAP} (assuming a minimal model).
A stronger constraint, $m_3 \circa{<} 0.1\eV$~\cite{epsP}, 
is obtained assuming thermal leptogenesis within  
see-saw models
with hierarchical right-handed neutrinos.
We reanalyzed this leptogenesis constraint,
merging the revised computation of dynamics of leptogenesis  
of~\cite{thermal}
with a revised  bound on the CP-asymmetry (see eq.\eq{epsDeg}).
It is weaker than previous bounds, and its validity needs extra  
assumptions
to discard a special (but non necessarily fine-tuned)
choice of parameters that can give a much larger asymmetry
(see eq.\eq{stimaDI} and fig.\fig{EpsDI}).
Furthermore in  appendix~\ref{fla}
we explained why  and which  single Boltzmann equation for the
total $B-L$ asymmetry is a good approximation
in the region where the constraint on neutrino masses is saturated.

From this revised analysis we obtain that if neutrinos
% If neutrinos will 
turn out to be lighter than 
\begin{equation}
m_i <0.15\eV
\end{equation}
thermal leptogenesis
can generically
produce the observed baryon abundancy.
This critical value is the present $3\sigma$ bound (see fig.\fig{m3})
and can mildly shift with more accurate measurements
of $\Delta m^2_{\rm atm}$, of $n_B$, or if supersymmetry will be  
discovered.

\medskip

We studied what happens dropping the dubious
assumption that hierarchical right-handed neutrinos give
quasi-degenerate neutrino masses.
If neutrinos are heavier than $0.15\eV$
quasi-degenerate $\nu_R$ would be suggested by good taste
and allow to weaken largely the leptogenesis constraint.
How much depends on why neutrinos are quasi-degenerate.
We consider two possible classes of interpretations:

\begin{itemize}
\item[a)] Neutrinos are not controlled by any flavour symmetry:
this naturally gives large mixing angles and
comparable neutrino masses, which might accidentally show
some mild level of quasi-degeneracy.
If we therefore assume $m_i\approx \tilde{m}_j$ we 
find that a mild degeneracy,  
$(M_2-M_1)/M_1 \sim 0.1$
is sufficient to push the leptogenesis constraint above 1 eV  
(continuous line in fig.\fig{mdeg}a).
This happens for two different reasons: the CP asymmetry can be  
resonantly enhanced
and is no longer suppressed by one power of the orthogonality
suppression 
factor  
%$\Delta m^2_{\rm atm}/[(m_3+m_1) \tilde{m}_j]\sim  A CHE SERVE?
$\Delta m^2_{\rm atm}/m_3^2 $, see eq.\eq{supprfactors}.
A sub-eV leptogenesis constraint would survive only if one of these two  
suppressions were present
(dashed line in fig.\fig{mdeg}a and previous analyses \cite{epsP}),
but this has no reason to generically occur.

\item[b)] Some flavour symmetry (e.g.\ SO(3)) keeps left and  
right-handed neutrinos quasi-degenerate giving
$\tilde{m}_j$ very close to neutrino masses $m_i$.
%       $1-m_i / m_j\approx 1-M_i/M_j\approx 1-\tilde{m}_i /  
%       \tilde{m}_j $  INESATTO, PERMETTE GRANDI mtildei
We find that this gives a CP-asymmetry
suppressed by 3/2 powers of the quasi degeneracy factor  
$1-m_1/\tilde{m}_j\sim \Delta m^2_{\rm atm}/m_3^2$, see eq.\eq{I2suppress},
resulting in a constraint $m_3 \circa{<} 0.6\eV$ (which can be  
largely relaxed if the $\tilde{m}_j$ are slightly larger than the $m_i$,
see fig.\fig{mdeg}b). 

\end{itemize}
In the last section we studied leptogenesis in alternative minimal  
models.
Neutrino masses can be mediated by tree-level exchange of
right-handed neutrinos, or of fermion SU(2)$_L$ triplets or of
scalar triplets.
We find that in the last two cases
leptogenesis can proceed enough out-of-equilibrium,
despite the new effect of gauge interactions.
The reason is that their rates are strongly Boltzmann suppressed
in the last stages of decay processes.
While fermion triplets lead to successful leptogenesis (giving
only slightly
stronger constraints than with singlets, eq.\eq{Tripletbound})
using only a single scalar triplet it seems impossible to achieve
a sufficiently large CP-asymmetry.

\paragraph{Acknowledgments}
This work has been partially supported by the EU under TMR
contracts HPRN--CT--2000--00148, HPRN--CT--2000--00152 and for
T.H.~by the EU
Marie Curie contract HPMF-CT-01765. We
thank Sacha Davidson for useful
suggestions and painful criticisms.

%%%%%%%%%%%%%%%%%%%%%%%%%%%%%%%%%%%%%%%%%%%%%%%%%%%%%%%%%%%%%%%%%%%%%%%%
\appendix

\section{Boltzmann equations with flavour}\label{fla}
We here explain how
the full network of Boltzmann equations can be approximated with a  
single equation
for the total $B-L$ asymmetry
when computing the constraint on quasi-degenerate neutrino masses.
In the standard approximation one writes one Boltzmann equation
for the total asymmetry, without caring about how
it is shared among different lepton doublet flavours.
In simple cases this can be a good
approximation~\cite{bcst} if done properly,
as the following example shows.
Let us suppose that $N_1$ decays generate a lepton asymmetry
in $\nu_1  = (\nu_\mu + \nu_\tau)/\sqrt{2}$
and that there are wash-out interactions acting
on $\nu_2 = (\nu_\mu - \nu_\tau)/\sqrt{2}$:
one can wonder if they are weighted by
a) $|\langle \nu_1 | \nu_{\mu,\tau}\rangle|^2 = 1/2$
or by
b) $|\langle \nu_1 | \nu_{2}\rangle|^2 = 0$?
The answer is a)
when scatterings induced by the $\tau$ Yukawa coupling
are much faster than the expansion of the universe
(because they convert $\nu_1$ into a incoherent mixture of $\nu_\mu$ and $\nu_\tau$)
and b) when they are much slower
(because $\nu_1$ remains a coherent superposition of $\nu_\mu$ and $\nu_\tau$).
Around the values $T\sim M_1 \gg 10^{11}\GeV$ for which
the leptogenesis constraint is saturated,
all SM lepton Yukawa couplings can be neglected (case b)
and the Boltzmann equation for leptogenesis is~\cite{bcst}
\begin{equation}\label{eq:rho}\frac{d\rho}{dt}=
zsH \frac{d\rho}{dz} =  \gamma_D(1-\frac{Y_{N_1}}{Y_{N_1}^{\rm eq}})
\frac{\Gamma \Pi - \bar\Gamma \bar \Pi}{\Gamma + \bar\Gamma}
-\frac{ \{\gamma_N,\rho\}}{8Y_L^{\rm eq}} \,,
\end{equation}
where $\rho$ is the $3\times 3$ matrix density that fully describes
how the 3 flavours share the $B-L$ asymmetry.
$\Pi$ ($\bar\Pi$) is the projector over the lepton (anti-lepton)
flavour to which $N_1$ decays with decay width $\Gamma$ ($\bar\Gamma$).
At tree level $\Gamma = \bar\Gamma = \Gamma_1/2$ and
$\Pi_{ij} =\bar\Pi_{ij} = ( \Pi_1)_{ij}\equiv
  \lambda_{1i}\lambda_{1j}^*/|\lambda\lambda^\dagger|_{11}$.
At one-loop $N_1$ decays into leptons and anti-leptons
with different rates 
(giving the total CP-asymmetry $\varepsilon_1 = (\Gamma-\bar\Gamma)/(\Gamma+\bar\Gamma)$)
and into different flavours ($\Pi \neq\bar\Pi$).
$\hat\gamma_N$ is the  $3\times 3$ flavour matrix
of interaction rates of $\Delta L=2$ scatterings.
It can be decomposed as $\hat\gamma_N = 4\gamma_D \Pi_1 +
\hat\gamma_N^{\rm sub}$,
where the first term takes into account resonant scatterings mediated
by on-shell $N_1$, and $\hat\gamma_N^{\rm sub}$ describes off-shell
scatterings mediated by $N_{1,2,3}$.
For all the other symbols we adopted the notations of \cite{thermal}
(e.g.\ $z= M_1/T$, $\gamma_D$ is the decay interaction rate,\ldots).

In general,
without making approximations
the matrix equation\eq{rho} cannot be replaced by a single equation
for the total asymmetry  $Y_{B-L} =  \hbox{Tr}\rho$,
nor by three equations for the diagonal components of $\rho$ (in
some flavour basis).
In the present case, taking into account that
  for quasi-degenerate neutrinos
  $\hat\gamma_N^{\rm sub}$ is a linear combination of
  $\Pi_1$ and of $1-\Pi_1$,\footnote{A less accurate approximation is
  $(\hat\gamma_N^{\rm sub})_{ij}\approx
\gamma_N^{\rm sub}\delta_{ij}$:
  when left-handed neutrinos are quasi-degenerate
  $\gamma_N^{\rm sub}$ is controlled by the average squared neutrino mass
  (rather than by their sum, which is 3 times larger).}
  it is non trivial to verify that at leading order in $\varepsilon_1$
  the solution to\eq{rho} is
$$ \rho = Y_{B-L} (\Pi_1+ \frac{\Pi-\bar\Pi}{2\varepsilon_1}) \,,$$
where $Y_{B-L}$ satisfies the Boltzmann equation of~\cite{thermal}
in `one flavour' approximation,
$$zsH\frac{dY_{B-L}}{dz} =
\varepsilon_1\gamma_D(1-\frac{Y_{N_1}}{Y_{N_1}^{\rm eq}})
-\frac{Y_{B-L}}{Y_L^{\rm eq}} (\frac{\gamma_D}{2} + 2\gamma_N^{\rm sub})
\,,$$
that is therefore adequate for studying
the heaviest neutrino mass compatible with leptogenesis.
This is the equation we used to calculate the constraint on neutrino masses.

\section{Maximal CP asymmetry with 2 quasi-degenerate $\nu_R$}\label{I2}
In the following we compute the maximum value of $I_2$ for fixed value 
of max$(\tilde{m}_1, \tilde{m}_2)$, which allows to give a bound on the
asymmetry both in the 2 quasi-degenerate case, eq.\eq{epsmaxdeg}, and
in the ``most realistic case'' of eq.\eq{epsreal}.
Using the parameterization in eq.\eq{seesawParamCI}, $I_2$ can be  
written as
\begin{equation}
I_2=\frac{1}{\tilde{m}_1 \, \tilde{m}_2 }
{\hbox{Im}[(R \cdot \hbox{diag}(m_1,m_2,m_3)\cdot   
R^{\dagger})_{12}^2]} \,,
\end{equation}
with
\begin{equation}\label{eq:constr}
\tilde m_1= (R \cdot  \hbox{diag}(m_1,m_2,m_3)\cdot R^{\dagger})_{11} \,,  
\qquad\hbox{and}\qquad
\tilde m_2 =(R\cdot  \hbox{diag}(m_1,m_2,m_3)\cdot  R^{\dagger})_{22} \,. 
\eeq
Neglecting the solar mass splitting, $I_2$ depends on 5 real  
parameters: the complex angles
$z_{13}$ and $z_{23}$ and the imaginary part of the angle $z_{12}$
(the real part of $z_{12}$ cancels out because $m_1=m_2$).
However, numerical inspection shows that the maximal value of $I_2$
can be obtained for any value of  $\hbox{Re} z_{13}$ and of $\hbox{Re}  
z_{23}$:
we can therefore put them to zero,
simplifying the expression and reducing the number of the free  
parameters to 3. Moreover,  for fixed
value of max$(\tilde{m}_1, \tilde{m}_2)$ the bound is obtained for
$\tilde{m}_1=\tilde{m}_2 \equiv \tilde{m}$. For fixed
$\tilde{m}_1$ the maximum of $\varepsilon_1+\varepsilon_2$ is
also obtained for $\tilde{m}_1=\tilde{m}_2$. 
All this together with eq.\eq{constr} gives   
{\small \begin{equation}\label{eq:asym2}
I_2^{\rm max}= 2 \max_z \,z \,\sqrt{1 - z} \bigg( 1 -  
\frac{m_1}{\tilde m} \bigg)
\bigg(1-\frac{m_1^2}{\tilde m^2}\bigg)^{-1/2}
\left(1 - \frac{\tilde m - {m_1}}{\tilde m + {m_1}}z \right)^{1/2}
\left(1 + \frac{\tilde m - {m_1}}{\tilde m + {m_3}} z\right)^{1/2}
\left(1 - \frac{\tilde m - {m_1}}{\tilde m + {m_3}}z\right)^{-1/2} ,
\end{equation}
}
where $z=[\cosh (2\, \hbox{Im}\,z_{23}) -1]( \tilde m + m_3 )/2( \tilde m - m_1) $
can  vary in the
interval $[0,1]$.
We now need to maximize the previous expression with respect to $z$.
This can be done analytically, but gives a lengthy expression.
In the limit $\tilde m \gg m_i$ the maximum is reached for $z =  
1/\sqrt{2}$,
while in the opposite limit $\tilde{m}\to m_1$ it is reached for $z = 2/3$.
Since these two values are  close (and since functions are almost flat  
around their maximum),
an excellent approximation is obtained by setting $z$ to any of these  
two numbers.
Assuming quasi-degenerate neutrinos, $m_1 \simeq m_3$, one finds a  
simple expression that interpolates between these two numbers,
%$z=[3-\sqrt{1+8r}]/4(1-\sqrt{r})$ (where ),
giving
\begin{equation}  I_2^{\rm max}  
=\sqrt{1+\frac{5}{2r}-\frac{1+(1+8r)^{3/2}}{8r^2}}
\approx (1-m_1/\tilde{m})^{3/2}
\label{eq:I2appendix}
\end{equation}
where $r=\tilde{m}^2/m_1^2$ and the last simple approximation is  
accurate to better than $10\%$.
So far we didn't put any restriction on $\tilde{m}_3$. Requiring in
addition
$\tilde{m}_3=\tilde{m}_{1,2}$ also gives eq.\eq{I2appendix}
with an accuracy better than $10\%$. We used therefore eq.\eq{I2appendix}
for all numerical analyses.
Note that $I_2^{\rm {max}}$ reaches unity asymptotically for large $\tilde{m}$.
 
%%%%%%%%%%%%%%%%%%%%%%%%%%%%%%%%%%%%%%%%%%%%%%%%%%%%%%%%%%%%%%%%%%%%%%%% 
%%%%
\footnotesize
\begin{multicols}{2}
  
\end{multicols}


\begin{thebibliography}{99}

\bibitem{FY}
M. Fukugita and T. Yanagida, Phys. Lett. {174B} (1986) 45.

\bibitem{leptogenesisBounds}
\hepart[hep-ph/0202239]{S. Davidson, A. Ibarra}.
The DI value of $\varepsilon_1$ was present also in
\hepart[hep-ph/0109030]{K. Hamaguchi, H. Murayama, T. Yanagida}.
An earlier work (\cite{bcst} around eq.~(5.1))
presented the DI bound as valid only up to cancellations,
in agreement with our present findings.



\bibitem{epsP}
\art[hep-ph/0302092]{W.~Buchmuller,
P.~Di Bari, M.~Pl\"umacher}{\NP}{B665}{445}{2003}.

\bibitem{thermal}
\hepart[hep-ph/0310123]{G.F. Giudice, A. Notari,
M. Raidal, A. Riotto, A. Strumia}

\bibitem{flanz} M. Flanz, E.A. Paschos, U. Sarkar, Phys. Lett.
{B345} (1995) 248;
M.~Flanz, E.A.~Paschos, U.~Sarkar, J.~Weiss,
Phys. Lett. {B389} (1996) 693; L. Covi, E. Roulet, F. Vissani, Phys. Lett. {B384}
(1996) 169; A. Pilaftsis, Phys. Rev. {D56} (1997) 5431; 
A. Pilaftsis, T.E.J. Underwood,  
hep-ph/0309342;
T.~Hambye, Nucl. Phys. {B633} (2002) 171.

\bibitem{pil2} A. Pilaftsis, Nucl. Phys. {B504} (1997) 61.

\bibitem{seesawsinglet}
  M. Gell-Mann, P. Ramond and R. Slansky,
   in {\it Supergravity}, edited by P. van Nieuwenhuizen and D. Freedman,
   (North-Holland, 1979), p.~315;
  S.L. Glashow, in Quarks and Leptons, Carg\`ese, eds. M. L\'evy et al.,
(Plenum, 1980, New-York), p. 707;
  T. Yanagida, in {\it Proceedings of the Workshop on the Unified Theory
   and the Baryon Number in the Universe}, edited by O. Sawada and
   A. Sugamoto (KEK Report No.~79-18, Tsukuba, 1979), p.~95;
  R.N.~Mohapatra and G. Senjanovi\'{c}, Phys. Rev. Lett. {44},
   (1980) 912.

\bibitem{tripletferm} R. Foot, H. Lew, X.-G. He and G.C. Joshi,
   Z. Phys. {C44} (1989) 441.
\bibitem{ma} E. Ma, Phys. Rev. Lett.  {81}
   (1998) 1171; E. Ma and D.P. Roy, Nucl. Phys. {B644} (2002) 290.

\bibitem{scalartriplet} G. Lazarides, Q. Shafi and C. Wetterich,
Nucl Phys. {B181} (1981) 287; R.N. Mohapatra and G. Senjanovi\'c, Phys.
Rev. {D23} (1981) 165; C. Wetterich, Nucl. Phys. {B187} (1981) 343.


\bibitem{seesawparam}
\hepart[hep-ph/0103065]{J.A. Casas, A. Ibarra}.


\bibitem{nucleo}
J. R. Ellis, J. Kim,  D. V. Nanopoulos,
Phys. Lett. B145, 181 (1984);
L. M. Krauss,
Nucl. Phys. B227, 556 (1983);
M. Yu. Khlopov, A. D. Linde,
Phys. Lett. 138B, 265 (1984);
J. R. Ellis, D. V. Nanopoulos, K. A. Olive, S.-J. Rey,
Astropart. Phys. 4, 371 (1996);
M. Bolz, A. Brandenburg, W. Buchmuller,
Nucl. Phys. B 606, 518 (2001).
R. H. Cyburt, J. Ellis, B. D. Fields, K. A. Olive,
Phys. Rev. D 67, 103521 (2003).
For a review see
M.Yu. Khlopov, `Cosmoparticle physics', World Scientific, 1999.

\bibitem{jarlskog}
\hepart[hep-ph/0312007]{S. Davidson, R. Kitano}.


\bibitem{bcst}
\art[hep-ph/9911315]{R.~Barbieri, P.~Creminelli,
A.~Strumia, N.~Tetradis}{\NP}{B575}{61}{2000}.


\bibitem{dipoli}
\art[hep-ph/9610485]{A. Romanino and A. Strumia}{\NP}{B490}{204}{1997}.

\bibitem{masar}
E. Ma and U. Sarkar, Phys. Rev. Lett. {80} (1998) 5716.

\bibitem{scaltriplepto} 
T. Hambye,
E. Ma and U. Sarkar, Nucl.
Phys. {B602} (2001) 23.

\bibitem{sod}  P. O'Donnell and
U. Sarkar, Phys. Rev. {D49} (1994)
2118.

\bibitem{hs} T. Hambye and G. Senjanovic, to appear in
Phys. Lett. {B}, hep-ph/0307237.

\bibitem{lazdent} See also G. Lazarides and Q. Shafi,
Phys. Rev. {D58} (1998) 071702. For a related model based on
inflation see also T. Dent, G. Lazarides and R. Ruiz de
Austri, hep-ph/0312033.



\bibitem{beta}
\art{The {\sc Mainz} collaboration}{Nucl.\ Phys.\ Proc.  
Suppl.}{91}{273}{2001};
\art{The {\sc Troitsk} collaboration}{Nucl.\ Phys.\ Proc.  
Suppl.}{91}{280}{2001}.



\bibitem{0n2b} The $0\nu2\beta$ bound on neutrino masses
is obtained  by combining $0\nu2\beta$ data, 
\art[hep-ph/0103062]{The {\sc Heidelberg--Moscow} collaboration,
H.V.~Klap\-dor-Klein\-grot\-haus {\it et al.}}{Eur.\ Phys.\  
J.}{A12}{147}{2001},
  with oscillation data: see
  \art[hep-ph/0201291]{F.~Feruglio, A.~Strumia and  
F.~Vissani}{\NP}{B637}{345}{2002}.
and references therein.



  \bibitem{WMAP}
The cosmological bound on neutrino masses
is obtained by combining WMAP data,
\hepart[astro-ph/0302207]{C. L. Bennett et al.},
with large-scale structure data: see
\hepart[astro-ph/0302209]{D. N. Spergel et al.}
and references therein.
This is similar to pre-WMAP analyses,
\art{A.~Lewis and S.~Bridle}{\PR}{D66}{103511}{2002}.
Weaker more conservative bounds are obtained in
\hepart[astro-ph/0303076]{S. Hannestad}{JCAP}{0305}{004}{2003}.




\end{thebibliography}
\end{document}